\documentclass[%
 reprint,
 amsmath,amssymb,
 aps
]{revtex4-2}

\usepackage{stackengine}
\usepackage{subcaption}
\usepackage[justification=raggedright,
            singlelinecheck=false]{caption}
\stackMath
\usepackage{graphicx}
\usepackage[
    starfontserif 
    ]{starfont}
\usepackage{multirow}
\usepackage{booktabs}
\usepackage{float}

\DeclareSymbolFont{starfontsym}{OT1}{sts}{m}{n}
\DeclareMathSymbol{\mathSun}{\mathord}{starfontsym}{115}
\DeclareMathSymbol{\mathMercury}{\mathord}{starfontsym}{102}
\DeclareMathSymbol{\mathVenus}{\mathord}{starfontsym}{103}
\DeclareMathSymbol{\mathTerra}{\mathord}{starfontsym}{76}
\DeclareMathSymbol{\mathvarTerra}{\mathord}{starfontsym}{108}
\DeclareMathSymbol{\mathMoon}{\mathord}{starfontsym}{100}
\DeclareMathSymbol{\mathvarMoon}{\mathord}{starfontsym}{97}
\DeclareMathSymbol{\mathMars}{\mathord}{starfontsym}{104}
\DeclareMathSymbol{\mathJupiter}{\mathord}{starfontsym}{106}
\DeclareMathSymbol{\mathSaturn}{\mathord}{starfontsym}{83}
\DeclareMathSymbol{\mathUranus}{\mathord}{starfontsym}{70}
\DeclareMathSymbol{\mathvarUranus}{\mathord}{starfontsym}{65}
\DeclareMathSymbol{\mathNeptune}{\mathord}{starfontsym}{71}
\DeclareMathSymbol{\mathPluto}{\mathord}{starfontsym}{74}
\DeclareMathSymbol{\mathvarPluto}{\mathord}{starfontsym}{72}

\usepackage{graphicx}
\usepackage{dcolumn}
\usepackage{bm,bbm,xcolor}
\usepackage[colorlinks=true, allcolors=blue]{hyperref}

\begin{document}

\preprint{APS/123-QED}

\title{Dark matter pair absorption}

\author{Martin Bauer}
 \affiliation{Institute for Particle Physics Phenomenology, Department of Physics, Durham University, Durham, DH1 3LE, United Kingdom}
\author{Javier Perez-Soler}%
\affiliation{Instituto de Física Corpuscular (CSIC-Universitat de València), Parc Científic UV, c/ Catedrático José Beltrán 2, E-46980 Paterna, Spain}%

\author{Jack D. Shergold}
\affiliation{Instituto de Física Corpuscular (CSIC-Universitat de València), Parc Científic UV, c/ Catedrático José Beltrán 2, E-46980 Paterna, Spain}

\date{\today}

\begin{abstract}
We present a comprehensive analysis of the sensitivity of atomic transitions to light dark matter pair absorption. Unlike scattering, where only a fraction of the dark matter energy is deposited, pair absorption processes absorb the full mass, and are therefore capable of constraining far lighter dark matter.  Spin-flip and fine structure transitions are able to constrain electroweak scale axial-vector couplings for bosonic dark matter with $\mu\mathrm{eV}$ to eV masses, whilst principal quantum number transitions are able to set similar constraints on bosonic dark matter with scalar couplings. Unfortunately, pair absorption is largely insensitive to light fermionic DM due its necessarily tiny density at low masses. We also demonstrate the sensitivity of pair absorption to the cosmic neutrino background, and find that spin-flip transitions can set constraints on the overdensity parameter of $\eta_\nu \lesssim 10^9$ for neutrino masses $m_\nu \lesssim 1\,\mathrm{meV}$, around a hundred times stronger than existing constraints.
\end{abstract}

\maketitle


\section{Introduction}\label{sec:introduction}
The detection of light dark matter (DM) is an incredible challenge due to the small momentum transfer of light, non-relativistic particles in scattering processes. Depending on the local DM density and mass, the rate for DM pair absorption can far exceed the scattering rate, and lead to a clean signal with fixed energy deposition which to leading order is independent of the DM velocity distribution. In addition, DM pair absorption can result in nuclear or atomic transitions that are heavily suppressed in the Standard Model of particle physics (SM). We present the first discussion of pair absorption here, focusing first on atomic transitions in hydrogen and hydrogenic ions, for which we can calculate precise transition rates using \texttt{CINCO}~\cite{Bauer:2024dbg}, and then extending our results to multi-electron atoms. 

Searches for new physics through absorption by nuclei have been a topic of discussion since the early studies of axions in the MeV mass range. These studies have already highlighted the advantages of utilizing forbidden magnetic monopole and magnetic dipole transitions~\cite{Donnelly:1978ty}, which have a clean signal with low photon backgrounds. Similarly, the absorption of fermionic DM by nuclei has been discussed in~\cite{Dror:2019dib}.

For atomic transitions,~\cite{Sikivie:2014lha, Vergados:2018qdb, Vergados:2024vky} proposed a technique where atoms are cooled into the ground state, and then undergo an axion-induced spin-flip transition to an excited state. A tuneable laser can then be used to drive excitations from the first to a second excited state, which in turn de-excites and emits signal photons. For spin-flip transitions, an external magnetic field can be used to scan over different axion masses. An experimental study for this detection scheme has shown that laser-induced backgrounds can be controlled~\cite{Braggio:2017oyt}.

Dark matter absorption for Si and Ge targets, and with an Al superconductor detector has been discussed in~\cite{Mitridate:2021ctr}. Additionally, spectroscopy using molecular transitions has been proposed~\cite{Arvanitaki:2017nhi}, as well as sensitivity studies for molecular oxygen transitions to the absorption of meV DM~\cite{Santamaria:2015gro}. Even smaller DM masses can be probed by measurements of the hyperfine splitting of hydrogen~\cite{Yang:2019xdz}, and the absorption and scattering of DM in material with spin-orbit couplings has been discussed in~\cite{Chen:2022pyd}. Finally, atomic transitions induced by spin-one vector bosons have been discussed in~\cite{Yang:2016zaz, Alvarez-Luna:2018jsb, Krnjaic:2023nxe,Graesser:2024cns}. 

In this paper, we demonstrate the unique sensitivity of spectroscopy to DM pair absorption. This process is possible for all DM quantum numbers, and can probe DM that only interacts quadratically, \textit{e.g.} if the DM candidate carries a global charge. We present a comprehensive analysis and identify the DM parameter space that can be probed using this approach, using the recently published code \texttt{CINCO} to numerically evaluate the overlap integrals and amplitudes~\cite{Bauer:2024dbg}. We then look at the sensitivity of spin-flip transitions to the relic neutrino pair absorption, and discuss the constraints that can be placed on the overdensity parameter.

\section{The advantages of pair absorption}\label{sec:advantages}
The scattering of DM on nuclei or electrons has long been the focus of direct detection programmes. This process is particularly relevant for massive, weakly-interacting DM, $\chi$, since it occurs even if all DM interactions with SM particles, $\psi$, are quadratic, \textit{e.g.} for operators of the form
\begin{equation}\label{eq:scat}
    \mathcal{L} = c_{\bar\chi\chi} \bar\chi \chi \bar \psi \psi,
\end{equation}
which are typical for stable DM candidates, and where $c_{\bar\chi \chi}$ denotes the coupling strength. For this process, the recoil energy satisfies $E_r \lesssim 2m_\psi m_\chi^2 \beta_\chi^2/(m_\chi + m_\psi)^2$, where $\beta_\chi \simeq 10^{-3}$ is the DM velocity~\footnote{We will work in units with $\hbar = c = k_B = 1$ throughout.}, $m_\psi$ is the SM fermion mass, and $m_\chi$ is the DM mass. For a given recoil energy threshold, the range of DM that can be probed via scattering is therefore limited. 

If instead, DM has linear interactions, \textit{e.g.} via the operator
\begin{equation}\label{eq:absorb}
\mathcal{L} = c_{\psi\phi}\,\phi \bar \psi \psi\,
\end{equation}
then it can be absorbed. Assuming a free target, and that such interactions do not change the SM species, the recoil energy is set by the DM mass, $E_r \simeq m_\phi$, independent of the DM velocity to leading order. As such, absorption processes are far more sensitive to lighter DM candidates~\cite{Ge:2022ius,Dzuba:2023zcq}. However, operators of the form~\eqref{eq:absorb} lead to DM decays into light SM species, such as neutrinos or photons, with a width proportional to $c_{\psi\phi}^2$, heavily constraining the interaction strength. 

There is one other intriguing possibility, however. Operators of the form~\eqref{eq:scat} give rise to DM \textit{pair absorption}, where the recoil energy is given by $E_r \simeq 2m_\chi$. Importantly, such operators do not induce DM decay, and so are far less constrained. As an additional advantage, pair absorption is possible for any DM quantum numbers. The cost is that the absorption rate is proportional to the square the DM number density, $n_\chi \simeq \rho_\mathrm{DM}/m_\chi$, which strongly suppresses interaction rates for intermediate to heavy DM. As we will see, however, this scaling is extremely advantageous for light DM, in particular light bosonic states. 

Here, we will focus on resonant, bound-bound atomic transitions induced by pair absorption. This differs slightly from the case presented thus far, as most of the energy deposited by the absorbed DM goes into exciting the atomic electron, rather than into the recoil energy of the atom. Nevertheless, the total energy deposited, $\Delta E_{fi} \simeq 2m_\chi$, remains far larger than that of a scattering process for the same DM mass, and can be observed \textit{e.g.} as a decay photon in a Sikivie-style experiment~\cite{Sikivie:2014lha}. We will first focus on hydrogen and hydrogenic ions as they are solvable systems with tractable calculations, and then use these results to estimate the transition rates for multi-electron atoms. From there, the extension to nuclei, semiconductors, and other gapped systems is straightforward. 

To illustrate our points, we will consider complex scalar dark matter $S$ with interactions described by an effective Hamiltonian that captures all operators at mass-dimension six,
\begin{equation}\label{eq:scalarHam}
    \begin{split}
        \mathcal{H}_S &=  m_e\left(S^\dagger S\right)\left(\bar \psi_e \left[G_{S,S} + iG_{S,P}\gamma^5\right] \psi_e\right) \\ 
        &+ i\left(S^\dagger \overleftrightarrow{\partial_\mu} S\right)\left(\bar \psi_e\gamma^\mu\left[G_{S,V} + G_{S,A}\gamma^5 \right] \psi_e\right),
    \end{split}
\end{equation}
where $\psi_e$ denotes the hydrogen electron field operator, and the $G_{S,i}$ are real, dimensionful couplings which scale with some new physics scale, $\Lambda$, as $\Lambda^{-2}$, and $m_e$ is the electron mass. The mass parameter of the complex scalar $S$ can in general receive weak-scale corrections, which are expected to remain small enough to allow the range of DM masses considered here. We will contrast this with fermionic dark matter, $\chi$, with the quadratic interaction terms
\begin{equation}\label{eq:fermionHam}
    \begin{split}
        \mathcal{H}_{\chi} &= \left(\bar \chi \gamma_\mu \left[G_{\chi,V_1} + G_{\chi,V_2}\gamma^5\right]\chi\right)\left(\bar \psi_e\gamma^\mu \psi_e\right) \\
        &+ \left(\bar \chi \gamma_\mu \left[G_{\chi,A_1} + G_{\chi,A_2}\gamma^5\right]\chi\right)\left(\bar \psi_e\gamma^\mu \gamma^5 \psi_e\right).
    \end{split}
\end{equation}
The simultaneous presence of CP-even and CP-odd couplings would in general lead to CP-violating observables. Throughout this work, however, we consider one operator at a time. We will specifically focus on the transitions for which only the scalar and axial-vector are relevant. As these transitions are, to leading order, forbidden for photons, they will have naturally low background rates due to photon absorption. On the other hand, pseudo-scalar and vector couplings share the same transitions, and so are far harder to constrain. These operators are also both CP-even, and so even if considered in tandem, would not lead to CP-violating effects.

\section{Dark matter pair absorption in Hydrogen}\label{sec:formalism}
We calculate atomic transitions in hydrogen induced by the pair absorption of DM using Fermi's golden rule
\begin{eqnarray}
    d\Gamma_\mathrm{p} = 2\pi |\mathcal{M}_{fi}|^2 \delta(\Delta E_{fi} - E - \bar E) \,d\rho_\mathrm{p},
\end{eqnarray}
where $E$ and $\bar E$ denote the energy of the absorbed particle and antiparticle, respectively, and the squared amplitude can, in general, be decomposed into the \textit{atomic tensor}, $A$, and expectation values of the absorbed field operators, $\mathcal{O}$, as~\cite{Bauer:2024dbg}
\begin{equation}\label{eq:ampSq}
    |\mathcal{M}_{fi}|^2 = \sum_{X,X}\langle\mathcal{O}_{X,\{\mu\}}\rangle\,\langle\mathcal{O}_{X',\{\nu\}}\rangle^*\, A_{X,X'}^{\{\mu\}\{\nu\}},
\end{equation}
with the indices $X,X' \in \{S,P,V,A,T\}$ running over the linearly-independent gamma matrix Lorentz structures present in $\mathcal{O}$, and where $\{\mu\}$ and $\{\nu\}$ denote the sets of Lorentz indices appearing in each. We will use \texttt{CINCO}~\cite{Bauer:2024dbg} to compute the squared amplitude, whose scale is set by the overlap integral of the initial and final state electron wavefunctions. For pair absorption, the phase space element is given by
\begin{equation}\label{eq:phaseSpaceElement}
    d\rho_\mathrm{p} = \frac{d^3p}{(2\pi)^3} \frac{d^3\bar{p}}{(2\pi)^3} g_\mathrm{DM}^2 f_\mathrm{DM}(p) f_\mathrm{DM}(\bar{p}),
\end{equation}
where $g_\mathrm{DM}$ is the number of DM degrees of freedom, and we assume that the DM and its charge conjugate share the same momentum distribution, $f_\mathrm{DM}$~\footnote{Note that~\eqref{eq:phaseSpaceElement} does not include the erroneous factors of the DM velocity that appear in the absorption phase space element given in~\cite{Bauer:2024dbg}.}. The momentum distribution of DM in the galactic reference frame can be written as
\begin{equation}\label{eq:fdark}
    f_{\mathrm{DM}}(p) =  \frac{\xi \rho_\mathrm{DM}}{2g_\text{DM} m_\mathrm{DM}^4  \beta_c^3} \exp\left(-\frac{p^2}{m_\mathrm{DM}^2\beta_c^2}\right),
\end{equation}
with $\beta_c = 7.8 \cdot 10^{-4}$ the circular velocity of the Milky Way, $\rho_\mathrm{DM} = 0.39\,\mathrm{GeV}\,\mathrm{cm}^{-3}$ the local DM density ~\cite{2023ApJ...946...73Z}, and $\xi \simeq 44.8$ a normalisation factor that accounts for the cutoff at the escape velocity, $\beta_\mathrm{esc} \simeq 5\beta_c/2$~\cite{escape_velocity}. There is one subtlety here. Fermions must obey the exclusion principle, such that the distribution function cannot exceed unity anywhere. The largest value of the distribution function occurs when $p = 0$, and so we require that
\begin{equation}
    \frac{\xi \rho_i}{2 g_\mathrm{DM} m_\mathrm{DM}^4 \beta_c^3} \leq 1,
\end{equation}
where $\rho_i$ is the energy density of the species under consideration. We can ensure that this condition is always enforced whilst maximising the density of the species that we are interested in by making the replacement
\begin{equation}\label{eq:TGeps}
    \rho_\mathrm{DM} \to \varepsilon \rho_\mathrm{DM}, \qquad \varepsilon = \min\left(1, \frac{2 g_\mathrm{DM}m_\mathrm{DM}^4 \beta_c^3}{\xi \rho_\mathrm{DM}}\right),
\end{equation}
for fermions. This is the Tremaine-Gunn bound~\cite{Tremaine:1979we}, which will drastically affect the sensitivity to light fermionic DM.

After transforming into the laboratory frame, we arrive at the pair absorption rate
\begin{equation}\label{eq:pairRate}
    \Gamma_\mathrm{p} = \frac{\xi^2}{128\pi^5} \frac{\rho_\mathrm{DM}^2}{ m_\mathrm{DM}^3 \beta_c^2} \mathcal{P}\big[ |\mathcal{M}_{fi}|^2\big],
\end{equation}
with the pair absorption phase space integral
\begin{equation}\label{eq:DM_P_integral}
    \begin{split}
    \mathcal{P}&\big[ |\mathcal{M}_{fi}|^2\big] =  \int d^2 \Omega_1\, d^2\Omega_2 \int_{0}^{x_+} dx\,|\mathcal{M}_{fi}|^2\\
    & \times \frac{x^2 y_\mathrm{p}^2}{y_\mathrm{p}+\omega_2} e^{-(x^2+y_\mathrm{p}^2)}\, \Theta\left(\frac{5}{2}-x\right) \Theta\left(\frac{5}{2}-y_\mathrm{p}\right).
  \end{split}
\end{equation}
Here, $\Theta(x)$ is the Heaviside step function, and $x$ and $y$ are rescaled velocity parameters. The remaining variables are defined by
\begin{align}
    x_+ &= -\omega_1 + \sqrt{\omega_1^2 + \omega_2^2 + 2(n_\mathrm{p}-1)},\\
    y_p &= -\omega_2 + \sqrt{ \omega_{2}^2-x^2 - 2 (x\omega_1 - n_\mathrm{p} + 1)},
\end{align}
and 
\begin{equation}
    \omega_i=\cos{\theta_\mathTerra} \cos\theta_i+\cos{\phi_i} \sin \theta_\mathTerra  \sin\theta_i,
\end{equation}
where $n_\mathrm{p} = (\Delta E_{fi} -2m_\mathrm{DM})/m_\mathrm{DM}\beta_c^2 \in [0,49/4]$ is a measure of the total kinetic energy carried by the incident DM pair, the angles $(\phi_i,\theta_i)$, $i\in \{1,2\}$ specify the direction of the incoming dark matter particle momenta relative to the quantization axis of the target atom, and $\theta_\mathTerra$ characterizes the orientation of Earth's motion through the dark matter halo; we set $\phi_\mathTerra=0$ without loss of generality. 

As we will make comparisons to scattering throughout, we also need the analogous rate. Following the same procedure, we find for the scattering rate
\begin{align}\label{eq:scatRate}
    \Gamma_\mathrm{s} = \frac{\xi \rho_\mathrm{DM}}{32\pi^5}\Big( m_\mathrm{DM}& \beta_c\mathcal{S}_0\left[|\mathcal{M}_{fi}|^2\right] \\
    &\pm \frac{\xi \rho_\mathrm{DM}}{2g_\mathrm{DM}m_\mathrm{DM}^3\beta_c^2}\mathcal{S}_Q\left[|\mathcal{M}_{fi}|^2\right]\Big),\notag
\end{align}
including the contributions from both particles and antiparticles, with $\mathcal{S}_0[f]$ and $\mathcal{S}_Q[f]$ the components of the phase space integral excluding and including quantum statistics, respectively, and where the + and - terms are taken, in turn, for bosons and fermions. The scattering phase space integrals are given by
\begin{align}
        \mathcal{S}_0&\left[|\mathcal{M}_{fi}|^2\right] =\int d^2 \Omega_1\, d^2\Omega_2 \int_{x_-}^{\infty} dx\, |\mathcal{M}_{fi}|^2 \\
        &\times \frac{x^2 y_\mathrm{s}^2}{y_\mathrm{s}+\omega_2} e^{-x^2}\,\Theta\left(\frac{5}{2}-x\right), \notag \\
    \mathcal{S}_Q&\left[|\mathcal{M}_{fi}|^2\right]= \int d^2 \Omega_1\, d^2\Omega_2 \int_{x_-}^{\infty} dx\, |\mathcal{M}_{fi}|^2 \label{eq:qs-int} \\
    &\times \frac{x^2 y_\mathrm{s}^2}{y_\mathrm{s}+\omega_2} e^{-(x^2+y_\mathrm{s}^2)}\, \Theta\left(\frac{5}{2}-x\right) \Theta\left(\frac{5}{2}-y_\mathrm{s}\right),\notag 
\end{align}
where for scattering
\begin{align}
    x_- &= -\omega_1 + \sqrt{\omega_1^2 - \omega_2^2 + 2n_\mathrm{s}}, \\
    y_\mathrm{s} &= -\omega_2 + \sqrt{\omega_{2}^2+x^2 - 2 (n_\mathrm{s} - x\omega_1)},
\end{align}
and $n_\mathrm{s} = \Delta E_{fi}/m_\mathrm{DM}\beta_c^2 \in [0,49/8]$. 

As the quantum statistical terms, $\mathcal{S}_Q$, are typically much smaller than the first term, $\mathcal{S}_0$, for the processes considered here, we will omit their full form and simply write $\pm\;\text{q.s.}$, where appropriate. Their contributions will be included in all plots and constraints. Additionally, when giving numeric values for the transition rates, we will use the maximum value of the phase space integrals across the $(n,\cos\theta_\mathTerra)$ plane. The finite width of these transitions, and its impact on experimental sensitivity is discussed at length in Appendix~\ref{app:pysr}, and taken into account when giving projections in Sections~\ref{sec:DMconstraints} and ~\ref{sec:cnub}.

We will consider three kinds of transitions: spin-flip transitions, for which only the magnetic quantum number, $m$, of the electron changes, fine structure transitions, for which only the total angular momentum, $j$, of the electron changes, and principal quantum number transitions, for which the principal quantum number, $n$, is also allowed to change. 

Spin-flip transitions have the advantage that the level splitting can be modified by the application of an external magnetic field, allowing us to scan over a range of DM masses~\cite{Sikivie:2014lha}. In hydrogen, this range is roughly $\mu\mathrm{eV}$ to $\mathrm{meV}$ via pair absorption. 

Fine structure and principal quantum number transitions have fixed transition energies determined by atomic structure, but do not require an external magnetic field. Hydrogen and members of its isoelectronic series can have tiny fine structure splittings, sub-$\mu\mathrm{eV}$ in some cases. Unfortunately, these have no fine structure transitions from their ground state, and would therefore require much more complicated experimental setups to search for excited state transitions. On the other hand, principal quantum number transitions are sensitive to $\mathrm{eV}$ scale DM and can always proceed from the ground state, and are uniquely sensitive to scalar vertices, \textit{e.g.} $G_{S,S}$ in~\eqref{eq:scalarHam}. Unfortunately, principal quantum number transitions suffer from much smaller overlap integrals than spin-flip and fine structure transitions. As such, we will consider both hydrogenic ions and multi-electron atoms for fine structure and principal quantum number processes; multi-electron atoms can have fine structure transitions from their ground state, typically in the $\mathrm{meV}$ to $\mathrm{eV}$ range, whilst their larger $Z_\mathrm{eff}$ results in more sizable overlap integrals for principal quantum number transitions, still maintaining $\mathcal{O}(\mathrm{eV})$ gaps where pair absorption is relevant.

\subsection{Spin-flip transitions: $1s_{1/2}^- \to 1s_{1/2}^+$}

The spin-flip transition $1s_{1/2}^- \to 1s_{1/2}^+$ is highly forbidden for photons, leading to a lifetime for the $1s_{1/2}^{+}$ level of $\tau(1s_{1/2}^-) \simeq 10^{7}\,\mathrm{y}$. On the other hand, particles with axial-vector couplings, \textit{e.g.} axions or neutrinos, can induce this transition at leading order~\cite{Bauer:2024dbg}. This combination of factors makes it an ideal candidate to search for new physics, as there is almost no background from photon scattering.

\begin{figure*}[t]
    \centering
    \includegraphics[width = \linewidth]{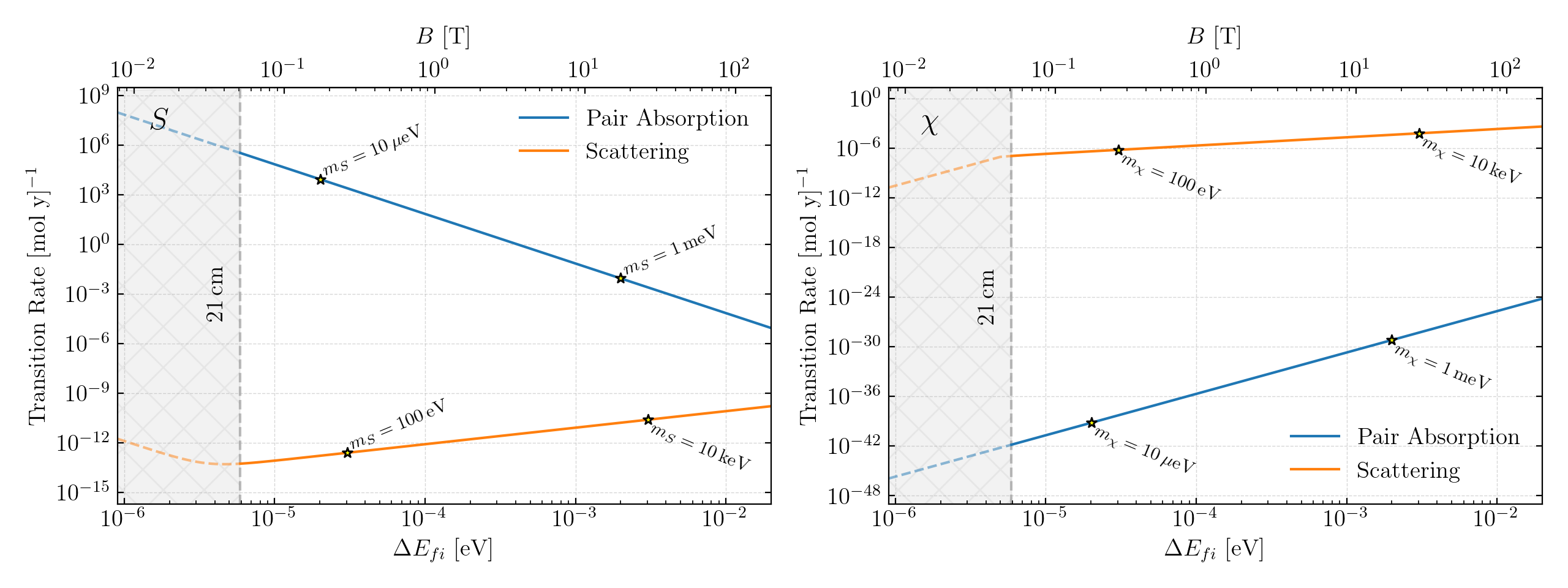}
    \caption{Left panel: scalar-induced, and right panel: fermion-induced $1s_{1/2}^- \to 1s_{1/2}^+$ transition rate in hydrogen, per mole of target atoms, for the pair absorption and scattering processes. The grey dashed line shows the energy of the $21\,\mathrm{cm}$ hydrogen hyperfine splitting, which effectively sets a lower bound on the size of the energy splitting that can be achieved using a magnetic field. The upper $x$-axis shows the approximate magnetic field strength required to induce the corresponding energy gap $\Delta E_{fi}$.}
    \label{fig:SpinFlipComparison}
\end{figure*}

The squared amplitude~\eqref{eq:ampSq} for this transition can in general be expressed as
\begin{align}\label{eq:spin-flipAmp}
    |\mathcal{M}_{fi}|^2 &= \left(\mathcal{I}_{ff} - \frac{1}{3} \mathcal{I}_{gg} \right)^2 \left|\langle\mathcal{O}_{A,x}\rangle -i\langle\mathcal{O}_{A,y}\rangle\right|^2
\end{align}
where the dimensionless radial overlap integrals are given, for example, by
\begin{equation}
    \mathcal{I}_{gf} = \int_0^\infty r^2dr\,g_{n',\kappa'}(r)f_{n,\kappa}(r),
\end{equation}
with $f_{n,\kappa}$ and $g_{n,\kappa}$ the upper and lower spinor components of the relevant hydrogen wavefunction, respectively, where the primes denote the final state, and $\kappa = (2j+1)(\ell-j)$, with $j$ and $\ell$ the total and orbital angular momentum quantum numbers of the state, respectively. Importantly, for the $1s_{1/2}^- \to 1s_{1/2}^+$ transition, the combination of $\mathcal{I}_{ff}$ and $\mathcal{I}_{gg}$ appearing in~\eqref{eq:spin-flipAmp} is almost exactly equal to one, with deviations at the $10^{-5}$ level. 

Considering the scalar Hamiltonian~\eqref{eq:scalarHam}, we therefore find the squared amplitude
\begin{equation}
    \left|\mathcal{M}_{fi}^S\right|^2 \simeq \frac{G_{S,A}^2 \beta_c^2}{4}(\vec{u}\mp \vec{v})_\perp^2,
\end{equation}
where $\vec{r}_\perp = \vec{r} - r_z \hat{e}_z$ denotes the component of a vector perpendicular to the polarisation of the target, whilst the upper sign is to be taken for pair absorption, and the lower for scattering. The variables $\vec{u}$ and $\vec{v}$ are the Galilean transformations of $\vec{x}$ and $\vec{y}$
\begin{align}
    \vec{u} = x \,\hat{n}_1 + \hat{n}_\mathTerra, \\
    \vec{v} = y \, \hat{n}_2 + \hat{n}_\mathTerra,
\end{align}
with $\hat{n}_i$ and $\hat{n}_\mathTerra$ the unit vectors colinear with the galaxy frame velocity of particle $i$, and the relative frame velocity, respectively. Note that $y$ should be replaced by either $y_\mathrm{p}$ or $y_\mathrm{s}$ for the pair absorption or scattering processes. 

Substituting our amplitudes into the pair absorption and scattering rates,~\eqref{eq:pairRate} and~\eqref{eq:scatRate}, respectively, and computing the phase space integrals following the steps in Appendix~\ref{app:pysr}, we find
\begin{align}
    \Gamma^S_\mathrm{p} &= \frac{G_{S,A}^2 \xi^2}{512\pi^5} \frac{\rho_\mathrm{DM}^2}{m_S^3}\mathcal{P}\left[(\vec{u} - \vec{v})_\perp^2\right] \\
    &=\left(\frac{8.61\cdot 10^{3}}{\mathrm{mol}\,\mathrm{y}}\,\right)\left[\frac{G_{S,A}}{G_F}\right]^2 \left[\frac{10\,\mu\mathrm{eV}}{m_S}\right]^3,\notag
\end{align}
which is roughly an $\mathcal{O}(\mathrm{mHz)}$ transition rate assuming electroweak scale couplings, $G_{S,A} \simeq G_F$. This transition is inaccessible to scalars with mass $m_S \lesssim 10\,\mu\mathrm{eV}$ due to the hyperfine splitting $\Delta E_{fi} \simeq 6\,\mu\mathrm{eV}$ at zero applied magnetic field. Note, however, that this specific cutoff only applies to the hydrogen $1s_{1/2}^- \to 1s_{1/2}^+$ transition; other hydrogen levels or different nuclei may have far smaller cutoffs, or in the case of scalar nuclei, no cutoff at all.

The scattering rate is given by
\begin{align}
    \Gamma_\mathrm{s}^S &= \frac{G_{S,A}^2\xi \rho_\mathrm{DM}}{128\pi^5}m_S \beta_c^3\mathcal{S}_0\left[(\vec{u}+ \vec{v})_\perp^2\right] +\text{q.s.} \\
    &= \left(\frac{2.45\cdot 10^{-14}}{\mathrm{mol\,y}}\right) \left[\frac{G_{S,A}}{G_F}\right]^2 \left[\frac{m_S}{10\,\mathrm{eV}}\right]+\text{q.s.},\notag
\end{align}
where the reference mass, $m_S = 10\,\mathrm{eV}$, represents approximately the same transition energy as $m_S = 10\,\mu\mathrm{eV}$ for the analogous pair absorption process, and we remind the reader that ``q.s.'' refers to the typically subleading quantum statistical term~\eqref{eq:qs-int}. We compare the scalar pair absorption and scattering rates as a function of the energy splitting in the left panel of Figure~\ref{fig:SpinFlipComparison}, and give a rough estimate of the external magnetic field required to produce the given gap. 

It should be immediately clear that the pair absorption process dominates over the scattering process, with a ratio as large as $10^{20}$ near the hyperfine splitting. This is largely due the quadratic DM density scaling of the absorption rate, which gives a huge rate enhancement at tiny DM masses. As the transition energy drops below $\Delta E_{fi} \simeq 4\,\mu\mathrm{eV}$, it can be seen that Bose-Einstein enhancement becomes relevant for scattering, giving the same quadratic scaling with density. As this is below the hyperfine splitting, however, it does not enhance the scattering rate in hydrogen, but may be relevant for other targets.

Following the same procedure for the fermionic Hamiltonian~\eqref{eq:fermionHam}, we arrive at the amplitude
\begin{align}\label{eq:fermion-spin-flip-amp}
    |\mathcal{M}_{fi}^\chi|^2 &= G_{\chi,A_i}^2\frac{\left(u-h_1 u_z\right) \left(v-h_2 v_z\right)}{u v}+\mathcal{O}(\beta_c)\\
    &= G_{\chi,A_i}^2 \,f(u,v) +\mathcal{O}(\beta_c), \notag
\end{align}
where $i = (1,2)$ for pair absorption, and scattering, respectively, whilst $h_1 = (h_\chi,h_{\chi,\mathrm{in}},-h_{\bar\chi,\mathrm{in}})$, $h_2 = (h_{\bar\chi},-h_{\chi,\mathrm{out}},h_{\bar\chi,\mathrm{out}})$ are the fermion helicities for pair absorption, particle, and antiparticle scattering, respectively. The terms proportional to $G_{\chi,A_{j}}$ with $j\neq i$ only appear at order $\beta_c$, and so any constraints on the associated coupling are at least three orders of magnitude weaker. For simplicity, we will consider the helicity averaged amplitude, in which case the pair absorption rate is given by
\begin{align}
    \Gamma^\chi_\mathrm{p} &\simeq \frac{G_{\chi,A_1}^2 \xi^2}{128\pi^5} \frac{\varepsilon^2\rho_\mathrm{DM}^2}{m_\chi^3\beta_c^2}\mathcal{P}[1] \\
    & =\left(\frac{6.20 \cdot 10^{-30}}{\mathrm{mol}\,\mathrm{y}}\right)\left[\frac{G_{\chi,A_1}}{G_F}\right]^2 \left[\frac{m_\chi}{1\,\mathrm{meV}}\right]^5,\notag
\end{align}
after taking into account the Tremaine-Gunn bound, assuming just two helicity degrees of freedom for the fermion, $g_\chi=2$. We give analytic approximations for both the helicity averaged and helicity dependent phase space integrals in Appendix~\ref{app:pysr}.

The analogous result for scattering is
\begin{align}
    \Gamma_\mathrm{s}^\chi &\simeq \frac{G_{\chi,A_2}^2\xi \,\varepsilon\rho_\mathrm{DM}}{16\pi^5}m_\chi \beta_c\mathcal{S}_0[1] -\text{q.s.}\\
     &= \left(\frac{5.99\cdot 10^{-6}}{\mathrm{mol}\,\mathrm{y}}\right) \left[\frac{G_{\chi,A_2}}{G_F}\right]^2\left[\frac{m_\chi}{1\,\mathrm{keV}}\right]-\text{q.s.},\notag
\end{align}
where we have set $\varepsilon = 1$ for our reference values, as the Tremaine-Gunn bound does not apply for the majority of the mass range relevant to scattering.

We show the rate of fermionic spin-flip transitions in the right panel of Figure~\ref{fig:SpinFlipComparison}. Clearly, spin-flip transitions are not well suited for fermionic DM detection. The fermionic pair absorption process only happens for masses at which the DM density must be tiny, a consequence of the Tremaine-Gunn bound. On the other hand, the masses involved in fermionic scattering are above the Tremaine-Gunn bound, but the scattering rate is nevertheless highly suppressed, since the local DM number density scales as $n_\chi \sim 1/m_\chi$, and their kinetic energy at these scales is too small to counteract this issue.

\subsection{Fine structure transitions: $n\ell_j \to n\ell_{j'}$}\label{sec:fs-transitions}

We now turn our attention to fine structure transitions, which we define as those which change only the total orbital angular momentum quantum number, $j$. We will begin by focusing on the $np_{1/2} \to np_{3/2}$ transition in hydrogen and hydrogenic ions, and then use these results to estimate the rates in multi-electron atoms, where the largest differences are the transition energies and numerical prefactors. 

\begin{figure*}[t]
    \centering
    \includegraphics[width = \linewidth]{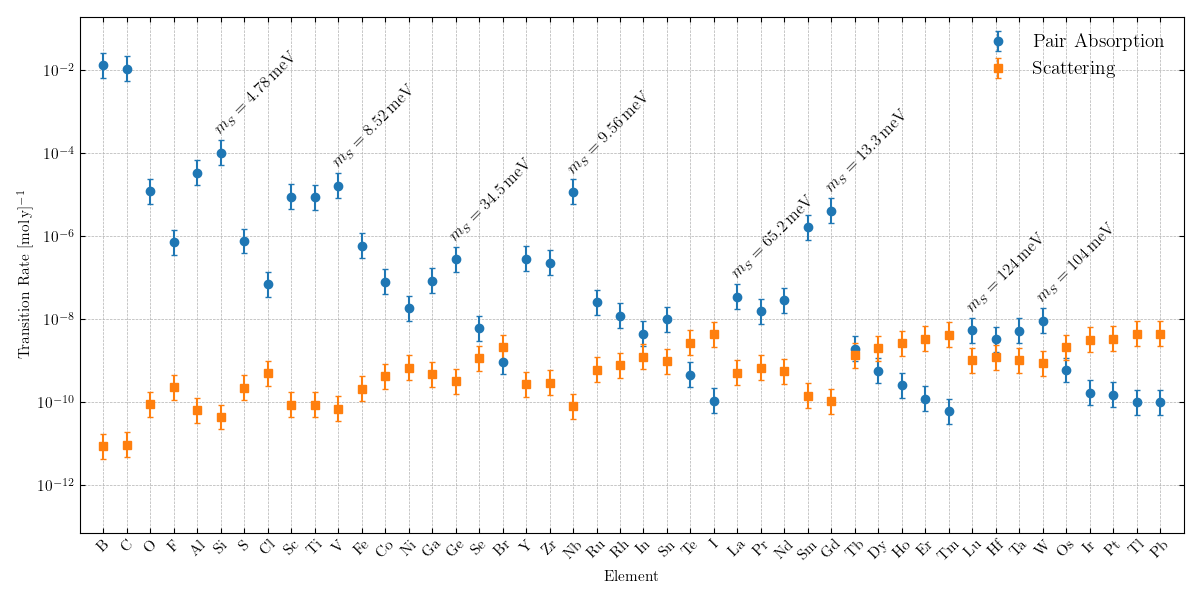}
    \caption{Scalar-induced, fine structure transition rates from the ground state in all stable, multi-electron elements, per mole of target atoms, for the pair absorption and scattering processes assuming $G_{S,A} = G_F$. We estimate the rates using the transition energies calculated using self-consistent field methods~\cite{NIST_ASD}, as given in Table~\ref{tab:fs_energies}, with a global uncertainty factor of two to account for any missed angular momentum coefficients and subleading configurations, if present.}
    \label{fig:fs_transitions_scalar}
\end{figure*}

Fine structure transitions differ significantly from the spin-flip transitions in that the splitting is fixed by atomic structure, and can not be controlled by an external magnetic field. In particular, the energy splitting between two fine structure levels with the same principal quantum number in hydrogenic ions is given by
\begin{equation}\label{eq:fs_splitting}
    \Delta E_\mathrm{FS} = \frac{2(\alpha Z)^4}{n^3}\frac{j_f - j_i }{(2j_i+1)(2j_f+1)} m_e + \mathcal{O}(\alpha^6 Z^6).
\end{equation}
There are two things to notice here. The first is the strong scaling with $Z\alpha$, which in theory could be used to probe a wide range of DM masses using highly charged ions, but in practice is infeasible due to large space charge effects. The second is the scaling with $n$, which leads to tiny gaps, potentially smaller than the hyperfine splitting, in highly excited hydrogen, especially for states with large angular momentum. Naturally, the use of excited states makes for a more challenging experiment with smaller target numbers than could be achieved with a ground state setup. However, the gains in the pair absorption rate, which scales as $\Delta E_{fi}^{-3} \sim n^{-9}$, along with the sensitivity to sub-$\mu\mathrm{eV}$ DM masses make this an appealing option.

The squared, polarisation-averaged amplitude for the $np_{1/2} \to np_{3/2}$ transition in hydrogen is given by
\begin{equation}\label{eq:fs_amplitude}
\left|\mathcal{M}_{fi}\right|^2 = \frac{8}{9} \left(\langle\vec{\mathcal{O}}_A\rangle\cdot\langle\vec{\mathcal{O}}_A\rangle^*\right)\,\mathcal{I}_{ff}^{2},
\end{equation}
with $\mathcal{I}_{ff}^2 \simeq 1$. As before, we consider the Hamiltonians~\eqref{eq:scalarHam} and~\eqref{eq:fermionHam}, for which the amplitudes are given by
\begin{align}
    \left|\mathcal{M}_{fi}^S\right|^2 &= \frac{2  G_{S,A}^2}{9} \beta_c^2 (\vec{u} \mp \vec{v})^2,\label{eq:amp-fs-s} \\
    \left|\mathcal{M}_{fi}^\chi\right|^2 &\simeq \frac{4 G_{\chi,A_i}^2}{9} \left(3 + h_1h_2\frac{\vec{u}\cdot \vec{v}}{uv}\right) +\mathcal{O}(\beta_c) \label{eq:amp-fs-f}\\
    &= \frac{4 G_{\chi,A_i}^2}{9} g(u,v) + \mathcal{O}(\beta_c)\notag,
\end{align}
for scalars and fermions, respectively, where the upper and lower signs are, in turn, for pair absorption and scattering. The remaining symbols are defined as in~\eqref{eq:fermion-spin-flip-amp}. For fermionic processes, the terms in $G_{\chi,j}$, $j \neq i$, enter at $\mathcal{O}(\beta_c)$, and so are velocity suppressed.

Plugging into the expressions for the rates, we find the pair absorption rate for scalars
\begin{align}\label{eq:fs_scalar_pair}
    \Gamma^S_\mathrm{p} &= \frac{G_{S,A}^2 \xi^2}{576\pi^5} \frac{\rho_\mathrm{DM}^2}{m_S^3}\mathcal{P}\left[(\vec{u} - \vec{v})^2\right] \\
    & =\left(\frac{988}{\mathrm{mol}\,\mathrm{y}}\right) \left[\frac{G_{S,A}}{G_F}\right]^2\left[\frac{n}{2}\right]^9,\notag
\end{align}
whilst for scattering we have
\begin{align}\label{eq:fs_scalar_scat}
    \Gamma_\mathrm{s}^S &= \frac{G_{S,A}^2\xi \rho_\mathrm{DM}}{144\pi^5}m_S \beta_c^3\mathcal{S}_0\left[(\vec{u} + \vec{v})^2\right] +\text{q.s.}\\
    &=\left(\frac{2.10\cdot 10^{-13}}{\mathrm{mol}\,\mathrm{y}}\right)\left[\frac{G_{S,A}}{G_F}\right]^2 \left[\frac{2}{n}\right]^3 + \text{q.s.},\notag
\end{align}
where the pair absorption mass is fixed to half the hydrogen $np_{1/2} \to np_{3/2}$ gap, which satisfies $\Delta E_{fi}= (45.3/n^3)\,\mu\mathrm{eV}$, and the scattering mass is fixed to $m_S = \Delta E_{fi}/\beta_c^2$. The tiny energy gaps in fine structure transitions once again strongly favour light scalar DM searches via pair absorption over intermediate masses via scattering. However, hydrogen and hydrogenic isotopes are otherwise unappealing candidates for fine structure transitions, as they must first be excited to \textit{e.g.} an $np_{1/2}$, $nd_{3/2}$, or $nf_{5/2}$ level before undergoing a transition, complicating any experimental efforts.

Following the same procedure for fermions, we find the helicity averaged rates
\begin{align}
    \Gamma^\chi_\mathrm{p} &\simeq \frac{G_{\chi,A_1}^2 \xi^2}{96\pi^5} \frac{\varepsilon^2\rho_\mathrm{DM}^2}{m_\chi^3\beta_c^2}\mathcal{P}[1] \\
    & =\left(\frac{4.94 \cdot 10^{-38}}{\mathrm{mol}\,\mathrm{y}}\right)
    \left[\frac{G_{\chi,A_1}}{G_F}\right]^2 \left[\frac{2}{n}\right]^{15},\notag
\end{align}
for pair absorption, once again assuming helicity is the only degree of freedom carried by the fermion. The analogous result for scattering is
\begin{align}
    \Gamma_\mathrm{s}^\chi &\simeq \frac{G_{\chi,A_2}^2\xi \,\varepsilon\rho_\mathrm{DM}}{12\pi^5}m_\chi \beta_c\mathcal{S}_0[1] -\text{q.s.} \\
    &= \left(\frac{5.95\cdot 10^{-7}}{\mathrm{mol}\,\mathrm{y}}\right)\left[\frac{G_{\chi,A_2}}{G_F}\right]^2\left[\frac{2}{n}\right]^{3}-\text{q.s.},\notag
\end{align}
with $\varepsilon = 1$ for our reference values. As with spin-flip transitions, the tiny energy gaps are far too small to detect any fermionic backgrounds. For pair absorption this is once again due to the Tremaine-Gunn bound, whilst for scattering this is due to the density scaling $1/m_\chi$ combined with the small kinetic energy of the incident DM. Unlike spin-flip transitions, this problem cannot be alleviated by increasing the gap size with a strong magnetic field to search for more energetic DM, and going to shells with higher $n$ only exacerbates the issue.

Many atoms with open $p$-, $d$-, or $f$-shells are able to undergo fine structure transitions from their ground state, and are a good alternative to hydrogenic ions. Their transition energies, typically $\mathcal{O}(\mathrm{meV})$, are still small enough for scalar pair absorption to be relevant, but large enough so that fermionic scattering may also be observable. Additionally, they completely eliminate the need to excite the targets, easing the experimental requirements significantly. 

We estimate the fine structure transition rate for all stable, multi-electron atoms in Figure~\ref{fig:fs_transitions_scalar}, using~\eqref{eq:fs_scalar_pair},~\eqref{eq:fs_scalar_scat}, and the energies tabulated in Table~\ref{tab:fs_energies}, taken from NIST~\cite{NIST_ASD}. For each, we modify the prefactor in the amplitude~\eqref{eq:fs_amplitude} from $8/9$ to $4/5$ for $d$-shell transitions, or $16/21$ for $f$-shells, and to account for the effects of electron screening, we use the $Z_\mathrm{eff}$ values from~\cite{LANZINI2015240} in the computation of $\mathcal{I}_{ff}$. We note, however, that in all cases for fine structure transitions, $\mathcal{I}_{ff} \simeq 1$ to a very good approximation, as the two fine structure levels in a given subshell have almost identical wavefunctions. Finally, we add a global uncertainty factor of two to the rate for each element, to account for any missed angular momentum coefficients and subleading configurations, which are at most $\mathcal{O}(1)$. 

\begin{figure*}[t]
    \centering
    \includegraphics[width = \linewidth]{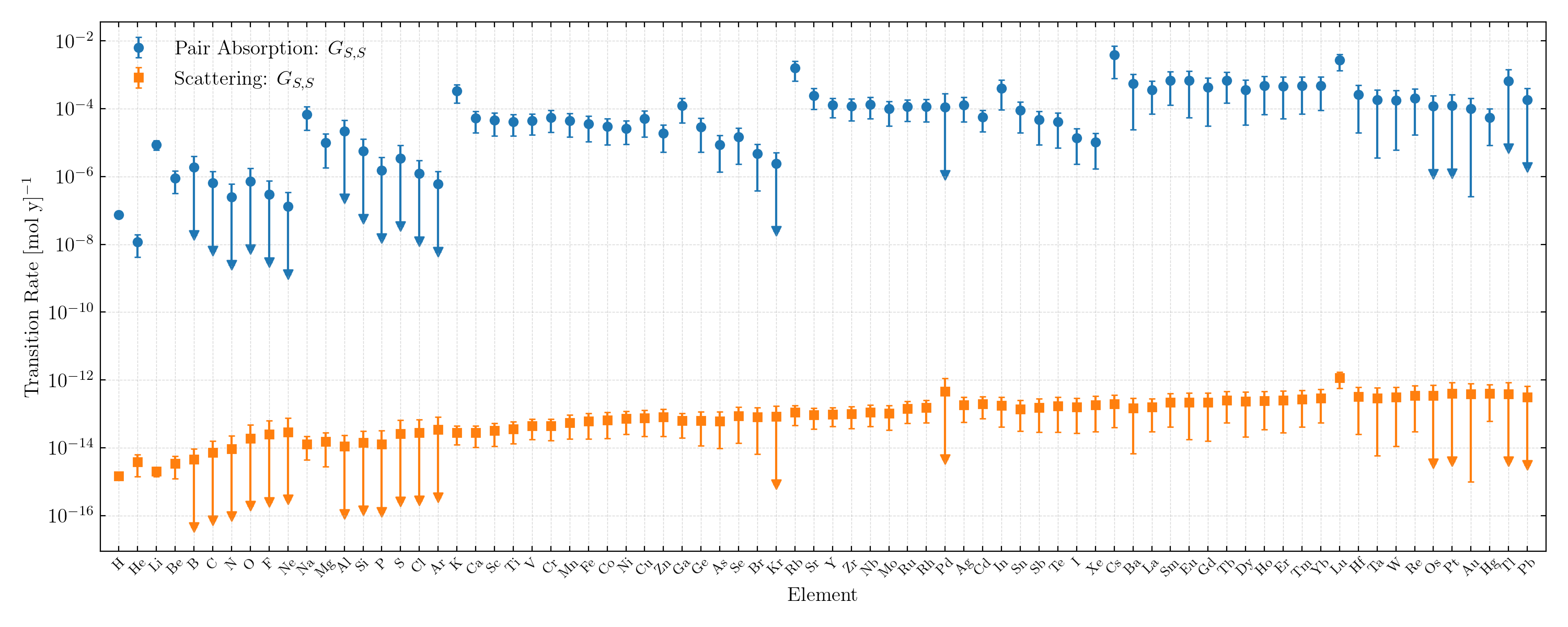}
    \caption{Scalar-induced, principal quantum number transition rates from the ground state in all stable, multi-electron atoms, per mole of target atoms, assuming $G_{S,S} = G_F$. We estimate the rates using the transition energies calculated using self-consistent field methods~\cite{NIST_ASD}, as given in Tables~\ref{tab:principal1} and~\ref{tab:principal2}. When the uncertainty on a point is as larger, or larger than the value, we cap the lower error bar with an arrowhead for readability. See the text for a discussion of the uncertainty calculation.}
    \label{fig:principal-rates}
\end{figure*}

Figure~\ref{fig:fs_transitions_scalar} demonstrates that scalar pair absorption favours light elements, with most capable of setting competitive constraints on meV scale DM with $\mathcal{O}(\mathrm{kg})$ of material. This is due to their smaller transition energies, which in general result from the $Z_\mathrm{eff}$ experienced by the transitioning electron. Several heavier elements, namely Nb, Sm, and Gd also stand out as good candidates for pair absorption. For all elements, the transition energies are too small for scattering to be relevant, requiring many tons of material for just a handful of events per year.  

\subsection{Principal quantum number transitions: $n\ell_{j}\to n'\ell_{j}$}\label{sec:pqn-transitions}
We finally focus on principal quantum number transitions, where all angular momentum quantum numbers remain constant and only the energy of the electron changes,  \textit{e.g.} $1s_{1/2} \to 2s_{1/2}$ in hydrogen. 

Principal quantum number transitions are typically much higher energy than spin-flip and fine structure transitions, with a transition energy given by the famous Rydberg formula
\begin{equation}
    \Delta E_n = \frac{m_e (\alpha Z)^2}{2} \left(\frac{1}{n_i^2} -\frac{1}{n_f^2}\right) + \mathcal{O}\left(\alpha^4 Z^4\right),
\end{equation}
to leading order in $\alpha Z$. As a result of their higher energy, these transitions are slightly less sensitive to pair absorption, but offer somewhat improved prospects for fermionic DM detection near the Tremaine-Gunn bound at $\mathcal{O}(10\,\mathrm{eV})$.

As before, we first focus on hydrogen, for which the squared, polarisation averaged amplitude for the $1s_{1/2} \to 2s_{1/2}$ transition is given by
\begin{align}\label{eq:amp-n}
    \left|\mathcal{M}_{fi}\right|^2&=\left(\langle\vec{\mathcal{O}}_A\rangle\cdot\langle\vec{\mathcal{O}}_A\rangle^*\right) \left(\mathcal{I}_{ff} - \frac{1}{3}\mathcal{I}_{gg}\right)^2 \\
    &+ \left(\mathcal{I}_{ff} - \mathcal{I}_{gg}\right)^2 \left|{\langle\mathcal{O}_S\rangle}\right|^{2},\notag
\end{align}
with $\mathcal{I}_{gg} \simeq -\mathcal{I}_{ff} = 5.5785\cdot 10^{-6}$. There are two things to notice from this amplitude: the first is that of those considered in this work, it is the only transition sensitive to the scalar coupling $G_{S,S}$~\footnote{At leading order in the long wavelength approximation. Other vertices may contribute at order $q^2\langle r\rangle^2 \ll 1$, with $q$ the modulus of the momentum transfer, and $\langle r\rangle$ the mean electron radius.}. The second is the smallness of the radial overlap integrals. This is a result of $\mathcal{I}_{ff}$, and consequently the entire transition rate, vanishing in the non-relativistic theory, leaving behind only the relativistic corrections at $\mathcal{O}(\alpha^2 Z^2)$. At the amplitude level, this scaling strongly favours transitions involving heavy hydrogenic ions, as $\mathcal{I}_{ff} \to 1$ as $Z \to 1/\alpha$, giving up to ten orders of magnitude compared to hydrogen. The rate also depends on the energy, however. For pair absorption, the rate scales as $m_\mathrm{DM}^{-3} \sim \Delta E_{fi}^{-3} \sim (Z\alpha)^{-6}$, and so lighter ions tend to be preferred despite their smaller overlap. On the other hand, for scattering, the the rate scales with $m_\mathrm{DM} \sim \Delta E_{fi} \sim (Z\alpha)^{2}$, strongly preferring heavier ions. For heavy multi-electron atoms, where the transitions of interest involve valence electrons with $n > 1$, the energies of the first principal quantum number transitions mostly lie in the $\mathcal{O}(1-10\,\mathrm{eV})$ range. Instead, the increased effective charge, $Z_\mathrm{eff}$, felt by the transitioning electron compensates for the smaller radial overlap between shells at larger radii, often leading to greater transition rates.

The amplitudes for the axial components of the amplitudes follow from direct comparison of~\eqref{eq:amp-n} with~\eqref{eq:fs_amplitude},~\eqref{eq:amp-fs-s}, and~\eqref{eq:amp-fs-f}, which including the scalar components yield the full amplitudes
\begin{align}
    \left|\mathcal{M}_{fi}^S\right|^2 &\simeq \frac{4  G_{S,A}^2}{9} \beta_c^2 (\vec{u} \mp \vec{v})^2 \mathcal{I}_{ff}^2 + \frac{G_{S,S}^2 m_e^2}{m_S^2} \mathcal{I}_{ff}^2,\label{eq:amp-principal-s} \\
    \left|\mathcal{M}_{fi}^\chi\right|^2 &\simeq \frac{8 G_{\chi,A_i}^2}{9} g(u,v)\mathcal{I}_{ff}^2, \label{eq:amp-principal-f}
\end{align}
with $g(u,v)$ and the index $i$ as given in~\eqref{eq:amp-fs-f}, and where we have used $\mathcal{I}_{gg} \simeq -\mathcal{I}_{ff}$. The scalar pair absorption is therefore
\begin{align}\label{eq:principal_scalar_pair}
    \Gamma^S_\mathrm{p} &\simeq \frac{ G_{S,A}^2 \xi ^2
  }{288 \pi ^5 }\frac{\rho _{\text{DM}}^2}{m_S^3} \mathcal{I}_{ff}^2\mathcal{P}[(\vec{u}-\vec{v})^2]\\
  &+\frac{G_{S,S}^2 \xi^2 }{128 \pi ^5 \beta _c^2}\frac{m_e^2 \rho _{\text{DM}}^2 }{m_S^5}\mathcal{I}_{ff}^2\mathcal{P}[1]\notag\\
    & =\left(\frac{5.38\cdot 10^{-24}}{\mathrm{mol}\,\mathrm{y}}\right)  \left[\frac{G_{S,A}}{G_F}\right]^2 \left[\frac{1}{Z}\right]^2\notag\\
    &+ \left(\frac{7.35\cdot 10^{-8}}{\mathrm{mol}\,\mathrm{y}}\right)  \left[\frac{G_{S,S}}{G_F}\right]^2 \left[\frac{1}{Z}\right]^6, \notag
\end{align}
where we take $m_S = \Delta E_{fi}/2$, with $\Delta E_{fi}/Z^2 = 10.2 \,\mathrm{eV}$ for the hydrogenic ion $1s_{1/2} \to 2s_{1/2}$ transition. The lack of velocity suppression, along with the $m_e/m_S$ enhancement factor, lead to a much larger absorption rate for the scalar coupling than the axial coupling for principal quantum number transitions. 

For scattering we instead find
\begin{align}\label{eq:principal_scalar_scattering}
    \Gamma^S_\mathrm{s} &\simeq \frac{G_{S,A}^2 \xi
    }{72 \pi ^5}\rho _{\text{DM}} m_S \beta_c^3 \mathcal{I}_{ff}^2 \mathcal{S}_0[(\vec{u}+\vec{v})^2]\\
   &+\frac{ G_{S,S}^2\xi}{32 \pi ^5}\frac{\rho_\mathrm{DM} m_e^2 \beta_c}{m_S}\mathcal{I}_{ff}^2\mathcal{S}_0[1] + \text{q.s.}\notag\\ 
    & =\left(\frac{2.94\cdot 10^{-18}}{\mathrm{mol}\,\mathrm{y}}\right)  \left[\frac{G_{S,A}}{G_F}\right]^2 \left[\frac{Z}{1}\right]^6\notag \\
    &+ \left(\frac{1.45\cdot 10^{-15}}{\mathrm{mol}\,\mathrm{y}}\right)  \left[\frac{G_{S,S}}{G_F}\right]^2 \left[\frac{Z}{1}\right]^2 + \text{q.s.},\notag
\end{align}
now with $m_S = \Delta E_{fi}/\beta_c^2$. Whilst the sensitivity to axial-vector couplings is slightly better than scalar pair absorption, and has a far more favourable scaling with $Z$, it is still far too weak to set any meaningful constraints, especially in the $m_S \sim \mathcal{O}(\mathrm{MeV})$ mass range. The sensitivity to scalar couplings is far worse, as the rate is now suppressed rather than enhanced by the $m_e/m_S$ factor.

Barring exotic models with a great many fermionic degrees of freedom, or highly modified DM distributions, principal quantum number transitions are more sensitive to fermionic pair absorption than spin-flip and fine structure transitions. Considering the helicity summed and averaged amplitude, we find the fermion pair absorption rate
\begin{align}\label{eq:principal_fermion_pair}
    \Gamma^\chi_\mathrm{p} &\simeq \frac{G_{\chi,A_1}^2 \xi^2}{48\pi^5} \frac{\varepsilon^2\rho_\mathrm{DM}^2}{m_\chi^3\beta_c^2}\mathcal{I}_{ff}^2\mathcal{P}[1] \\
    &=\left(\frac{1.78 \cdot 10^{-21}}{\mathrm{mol}\,\mathrm{y}}\right)
    \left[\frac{G_{\chi,A_1}}{G_F}\right]^2 \left[\frac{1}{Z}\right]^{2} \left[\frac{\varepsilon}{9.57\cdot 10^{-3}}\right],\notag
\end{align}
accounting for the Tremaine-Gunn bound, assuming $g_\chi = 2$. Unlike the other transitions, which are either heavily suppressed by (pair absorption) or lie firmly above (scattering) the Tremaine-Gunn bound, principal quantum number transitions via fermionic pair absorption lie on the border. In hydrogen, assuming the distribution function~\eqref{eq:fdark}, the suppression factor is $\varepsilon \simeq 10^{-2}$, and is evaded completely when $g_\chi \gtrsim 209$. The mass of the absorbed DM also exceeds the bound for $Z \geq 2$ in hydrogenic ions, and so there is no suppression there either. For these reasons, principal quantum number transitions are the most sensitive to fermion pair absorption. In spite of this, the expected rates are still extremely small, and we can only place very weak constraints on the axial-coupling $G_{\chi,A_1}$.

The fermionic scattering rate is given by
\begin{align}\label{eq:principal_fermion_scattering}
    \Gamma_\mathrm{s}^\chi &\simeq\frac{G_{\chi,A_2}^2\xi \,\varepsilon\rho_\mathrm{DM}}{6\pi^5}\mathcal{I}_{ff}^2m_\chi \beta_c\mathcal{S}_0[1] - \text{q.s.} \\
    &= \left(\frac{8.31\cdot 10^{-12}}{\mathrm{mol}\,\mathrm{y}}\right) \left[\frac{G_{\chi,A_2}}{G_F}\right]^2\left[\frac{Z}{1}\right]^6 -\text{q.s.},\notag
\end{align}
far greater than the pair absorption rate due to the larger masses, $\mathcal{O}(\mathrm{MeV})$ involved in principal quantum number transitions. 

As before, we now turn our attention to principal quantum number transitions in multi-electron atoms. These are significantly more difficult to estimate than fine structure transitions, as in general, the radial overlap integrals are much smaller than one, and depend heavily on the effective charge of the transitioning electron. We estimate the effective charge of the transitioning electron, $Z_\mathrm{eff}$, as the arithmetic mean of its effective charge in the initial and final state configurations, $Z_i$ and $Z_f$, respectively, with an uncertainty of half the range. We then use \texttt{CINCO} to estimate the value of $\mathcal{I}_{ff} \simeq -\mathcal{I}_{gg}$ and its uncertainty. Finally, we combine this with the transition energies computed using self-consistent field methods as given by NIST~\cite{NIST_ASD} to estimate the rate for each multi-electron atom. For each element, we only consider the first excited state with the same angular momentum quantum numbers. Higher $\Delta n$ transitions may also be of interest to probe DM of different masses, but in general will have smaller rates due to the larger transition energies and smaller overlap integrals.

The values of $\Delta E_{fi}$, $Z_\mathrm{eff}$, and $\mathcal{I}_{ff}$ for each transition are given in Tables~\ref{tab:principal1} and~\ref{tab:principal2}, whilst the transition rates for the scalar coupling $G_{S,S}$ are shown in Figure~\ref{fig:principal-rates}. We do not show any of the rates for the axial couplings, as they are significantly smaller and better constrained using spin-flip and fine structure transitions. Clear trends can be seen across each period. The alkali metals (Li, Na, K, Rb, and Cs) have the largest rates within each period, as their first excited states have the lowest transition energies. They also feature the smallest uncertainties as they are the most hydrogen-like elements, and so the transitioning electron is well described by a hydrogenic wavefunction. 
Some of these elements, namely Rb and Cs, are already used to constrain DM-SM interactions~\cite{Hees:2016gop,Tino:2020dsl}, albeit using far fewer than a mole of atoms. Due to their higher transition energy, however, there is a reduced need to cool the target for principal quantum number transitions, which may allow for significantly higher numbers of atoms.

\section{Coherence}\label{sec:coherence}

Coherent elastic scattering can lead to an enhanced scattering rate that scales as the square of the number of target atoms, $\Gamma \sim N_T^2$, provided the momentum transfer satisfies $qR < 1$, where $R$ is the size of the target. When this condition is not met, the rate scales linearly, $\Gamma \sim N_T$. Coherent absorption and inelastic scattering processes require the target atoms to be in quantum superpositions of ground and excited states. The maximal enhancement occurs when the system is prepared in a \emph{Dicke state}, a highly entangled state in which a single excitation is delocalized across many atoms~\cite{Arvanitaki:2024taq}. These states enable the ensemble to absorb energy collectively, analogous to the collective response seen in coherent elastic scattering. Such superpositions have been experimentally realized in systems with hundreds of atoms embedded in crystals~\cite{Zarkeshian2017}. However, realistic targets would necessitate macroscopic ensembles containing more than $N_T > \sqrt{N_A} \approx 10^{12}$ atoms, where $N_A$ is Avogadro's number. 

\begin{figure*}[t]
    \centering
    \includegraphics[width = \linewidth]{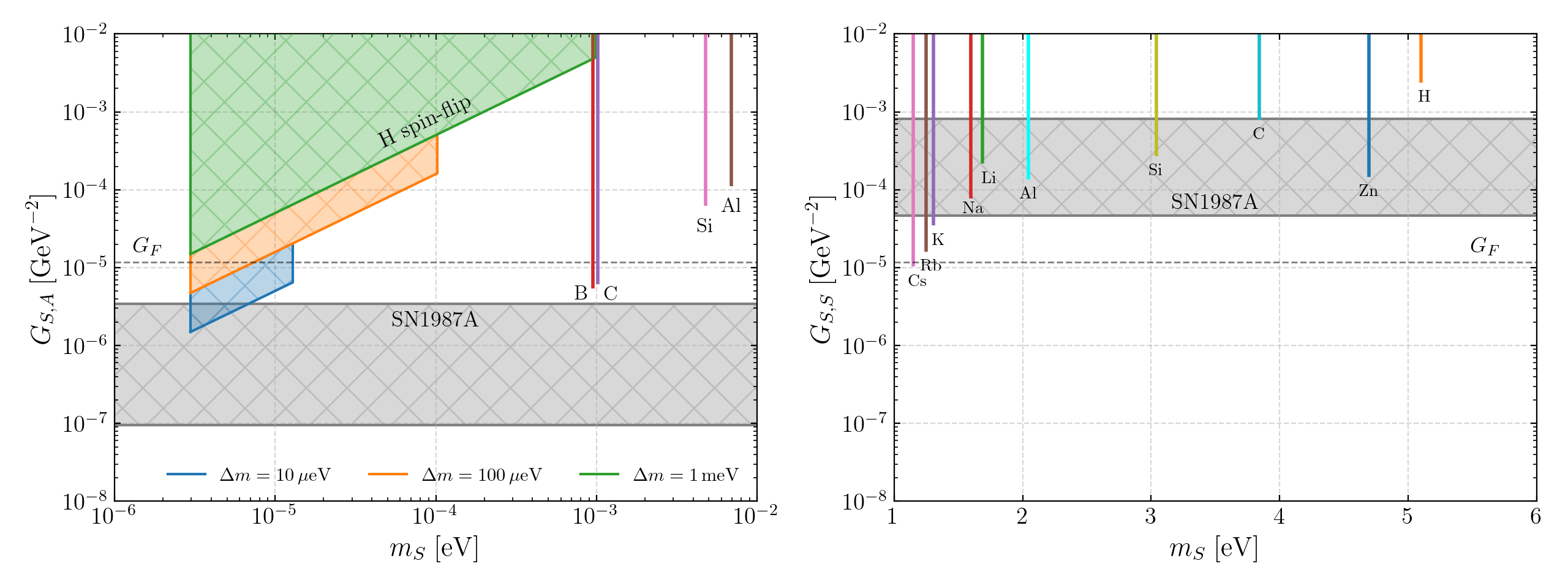}
    \caption{Sensitivity to dark matter pair absorption via the axial coupling $G_{S,A}$ (left panel), the scalar coupling $G_{S,S}$ (right panel). The filled regions correspond to a hydrogen spin-flip transition, and assume a scan range of $\Delta m = 10\,\mu\mathrm{eV}$ (bottom), $100\,\mu\mathrm{eV}$ (middle), and $1\,\mathrm{meV}$ (top). Left panel: From left to right, the vertical lines are for fine structure transitions in B, C, Si and Al. Right panel: From left to right, the vertical lines are for principal quantum number transitions in Cs, Rb, K, Na, Li, Al, Si, C, Zn, and H.  All curves assume $10^3$ moles of target, an observation time of one year, and do not assume any coherence as discussed in Section~\ref{sec:coherence}.}
    \label{fig:PASensitivity}
\end{figure*}

\section{Sensitivity of experimental searches}\label{sec:DMconstraints}
The homogeneous evolution of a complex scalar dark matter with masses of the order of eV or below can be described by solving the Klein-Gordon equation in a Friedmann-Lemaitre-Robertson-Walker geometry~\cite{Yang:2025vcb}. The late-time behaviour of both the real and the imaginary parts of the complex scalar field $S(x)=\varphi(x)+iA(x)$ oscillate with a frequency set by their mass\footnote{Aspects of the quantum descriptions of dark matter in the sub-eV mass regime are discussed in D. Y. Cheong, N. L. Rodd, and L.-T. Wang, \href{https://journals.aps.org/prd/abstract/10.1103/PhysRevD.111.015028}{Phys. Rev.
D 111, 015028 (2025),} \href{https://arxiv.org/abs/2408.04696}{arXiv:2408.04696 [hep-ph]}}, such that
\begin{align}\label{eq:osc}
S(x)= \varphi_0 \cos(m_S t) +i A_0 \sin(m_S t)\,, 
\end{align}
where we neglect phases suppressed by the DM velocity. For real scalars, these oscillations can induce stringent constraints from time-varying fundamental constants, such as the fine-structure constant or the electron mass. In the case of a complex field with interactions of the form~\eqref{eq:scalarHam}, both the real and imaginary components of \eqref{eq:osc} oscillate. However, if the real and imaginary parts have equal mass, so that  $\varphi_0=A_0 =\sqrt{2\rho_\text{DM}}/m_S$, the time-dependent contributions to the electron mass cancel to leading order, resulting in no oscillations. In addition, the pair absorption processes discussed here have a relevant time-scale set by the width of the atomic transition, which is many orders of magnitude larger than the dark matter oscillation frequency for the relevant parameter space.

A second important effect of ultralight dark matter with quadratic interactions is an effective, matter-dependent mass that induces non-trivial solutions to the equations of motion near massive bodies, such as planets~\cite{Hook:2017psm, Hees:2018fpg, Beadle:2023flm,  Bauer:2024yow, Bauer:2024hfv, Banerjee:2025dlo, delCastillo:2025rbr}. The gradient of the dark matter field can then induce an acceleration that can be probed in tests of the equivalence principle~\cite{Banerjee:2022sqg}. In the case of scalar fields with quadratic, scalar interactions with electrons, the strongest limits have been set by the MICROSCOPE satellite~\cite{Touboul:2017grn, Touboul:2022yrw}. We emphasize, however, that this constraint does not apply to other coupling structures with spin-dependent couplings, \textit{e.g.} $G_{S,P}$ and $G_{S,A}$, in the low-energy limit, as the test mass is unpolarized. 

Light, quadratically-coupled dark matter can also play a key role in supernova evolution~\cite{Olive:2007aj,Cox:2024oeb,Manzari:2023gkt,Guha:2018mli} through pair annihilation and scattering processes, $e^+e^- \to S^*S$ and $e^\pm S^{(*)} \to e^\pm S^{(*)}$, respectively. However, unlike the MICROSCOPE bound, these apply to both spin-dependent and spin-independent couplings. Following the formalism of~\cite{Guha:2018mli}, we find the excluded regions $4.6 \cdot 10^{-5}\,\mathrm{GeV}^{-2} \leq G_{S,S} \leq 8.1 \cdot 10^{-4}\,\mathrm{GeV}^{-2}$, and $9.6 \cdot 10^{-8}\,\mathrm{GeV}^{-2} \leq G_{S,A} \leq 3.5 \cdot 10^{-6}\,\mathrm{GeV}^{-2}$.

In Figure~\ref{fig:PASensitivity} we compare these limits with the projected sensitivity of different targets to pair absorption. In the case of spin-flip transitions we follow~\cite{Yang:2019xdz} and consider a target of $N_T=10^3 \,\mathrm{mol}$ of hydrogen. red The experimental realisation of such a target is very challenging and depends on many factors. Cosmic ray interactions and thermal noise require excellent shielding and very low temperatures $<20$ mK depending on the dark matter mass. For spin-flip transitions, the signal could be detected either by Stern–Gerlach-type experiments, separating and counting atoms in the excited spin state, or by measuring the resulting magnetic signal in a maser-like storage bulb~\cite{Yang:2019xdz}.

The effective observation time of the search is reduced by an amount dependent on the mass range explored, $\Delta m_\mathrm{DM}$
\begin{equation}
    t_\mathrm{eff} = \mathrm{min}\left(1,\frac{\Delta n\, m_\mathrm{DM} \beta_c^2}{\Delta m_\mathrm{DM}}\right) t_\mathrm{exp}.
\end{equation}
The width of each phase space integral, $\Delta n$, is given in Table~\ref{tab:pysr-fits}, alongside details of their derivation in Appendix~\ref{app:pysr}.

We assume $t_\text{exp}=1$ year and require 3 signal events, corresponding to a $\sim 91\%$ confidence limit signal in the absence of background
\begin{align}
N_T\Gamma_p^S t_\text{int}>3,
\end{align}
and show the parameter space for scans over the ranges $\Delta m_\mathrm{DM} = 10\,\mu\mathrm{eV}$, $100\,\mu\mathrm{eV}$, and $1\,\mathrm{meV}$ in the left panel of Figure~\ref{fig:PASensitivity}. For the narrow scan range, the sensitivity of spin-flip transitions to $G_{S,A}$ is always sufficient to constrain electroweak scale couplings, $G_{S,A} \simeq G_F$, and near the hyperfine splitting, is able to set constraints complimentary to those from SN1987A. Wider band searches, $\Delta m_\mathrm{DM} = 10, 100\,\mathrm{meV}$, are able to probe electroweak scale couplings near the hyperfine splitting, weakening to weakening to $G_{S,A} \simeq 10^{3}\,G_F$ for $m_S \simeq 1\,\mathrm{meV}$. Importantly, spin-flip transitions are capable of constraining new parameter space, regardless of the scan range chosen.

For both fine-structure transitions and principal quantum number transitions the transition energy is independent of external magnetic fields and probes an extremely narrow range of dark matter masses. We show the projected sensitivity for fine-structure transitions induced by scalar pair absorption in the left panel of Figure~\ref{fig:PASensitivity} with vertical lines, for (from left to right) B, C, Si, and Al. Here, we assume the same number of targets, $N_T=10^3\,\mathrm{mol}$, a runtime $t_\mathrm{exp}=1\,\mathrm{y}$, and require 3 signal events, such that 
\begin{align}
N_T\Gamma_p^S t_\text{exp}>3.
\end{align}
These are much more sensitive than the spin-flip transitions at a given mass, as the effective exposure time is not `diluted' by having to scan over multiple masses. Instead, we are limited by only being able to constrain a very narrow mass band, $\Delta m_S \simeq 10^{-6}\, m_S$ around the central transition energy. 

Principal quantum number transitions are very similar in this regard, but sit at much higher energies than fine-structure transitions. They are also the only transitions considered here that are sensitive to the scalar coupling $G_{S,S}$. We show the projected sensitivity for principal quantum number transitions induced by scalar pair absorption in the right panel of Figure~\ref{fig:PASensitivity} with vertical lines, for (from left to right) Cs, Rb, K, Na, Li, Al, Si, C, Zn, and H. The alkali metals are able to set the strongest constraints, on the scalar coupling $G_{S,S}$, all in the mass range 1-2 eV. Of these, Rb, Cs and K are capable of probing electroweak scale couplings, $G_{S,S} \simeq G_F$, whilst the remaining elements are able to set constraints one or two orders of magnitude weaker. The heavier alkali metals are also capable of constraining couplings weaker than the free-streaming supernova bound, $G_{S,S} \simeq 4.6 \cdot10^{-5}\,\mathrm{GeV}^{-2}$. Transitions between principal quantum numbers can produce optical de-excitation photons that are directly detectable, thereby avoiding many of the challenges associated with hyperfine-structure transitions. As an alternative, the excited state may be read out using laser-assisted de-excitation~\cite{Sikivie:2014lha}.

\section{Neutrino pair absorption}\label{sec:cnub}
The fermion Hamiltonian~\eqref{eq:fermionHam} is equivalent to the low energy effective Hamiltonian for neutrino-electron weak interactions. To recover the SM interaction, one simply sets $\chi \to \nu$, followed by $G_{\chi,V_1} = -G_{\chi,V_2} \to G_F g_V/\sqrt{2}$, and $G_{\chi,A_1} = -G_{\chi,A_2} \to -G_F g_A/\sqrt{2}$, with $g_V = -1/2 + 2\sin^2\theta_W$ and $g_A = -1/2$ the vector and axial couplings, respectively, given in terms of the Weinberg angle $\theta_W$. 

An interesting question then arises: \textit{can we detect neutrinos using pair absorption?} In particular, we want to focus on the constraints that can be placed on the relic neutrino density using this method, which is largely unconstrained experimentally~\cite{Bauer:2022lri}. Of the transitions considered here, spin-flip transitions seem best suited to this task, as they allow us to scan over the range of possible neutrino masses.

Following from~\eqref{eq:spin-flipAmp}, the squared amplitude for relic neutrino pair absorption is
\begin{equation}
    \left|\mathcal{M}_{fi}\right|^2 = \frac{G_F^2}{8}\frac{(E_\nu + p_\nu)(E_{\bar\nu} + p_{\bar\nu})}{E_\nu E_{\bar\nu}}(1-c_\nu c_{\bar\nu}),
\end{equation}
where $c_\nu$ and $c_{\bar\nu}$ are the cosines of the polar angles between the axis of quantisation and the incident neutrino and antineutrino, respectively. Note that as we do not know for certain whether relic neutrinos are relativistic today, we do not make any non-relativistic expansions. We have also set $h_\nu = -1$, and $h_{\bar\nu} = +1$, assuming only left helicity neutrinos and right helicity antineutrinos. 

Relic neutrinos are expected to follow a massless Fermi-Dirac distribution today, at a temperature $T_{\nu,0} \simeq 0.168\,\mathrm{meV}$
\begin{equation}
    f_\mathrm{FD}\left(p\right) = \frac{1}{\displaystyle\exp\left(\frac{p}{T_{\nu,0}}\right) +1},
\end{equation}
which modifies the phase space integral~\eqref{eq:DM_P_integral}. Fortunately, as the neutrino distribution is isotropic, the angular part of the is far easier to compute, and we are left with the absorption rate
\begin{equation}\label{eq:cnubRate}
\Gamma_{\mathrm{C}\nu\mathrm{B}} = \frac{G_F^2 \pi}{9 \zeta (3)^2}\frac{n_{\nu,0}^2}{T_{\nu,0}} \mathcal{P}_\nu,
\end{equation}
with $n_{\nu,0} \simeq 56\,\mathrm{cm}^{-3}$ the predicted relic neutrino density per degree of freedom, $\zeta$ the Riemann zeta function, and $\mathcal{P}_\nu$ the dimensionless neutrino phase space integral
\begin{equation}
    \mathcal{P}_\nu = \int_0^{x^+_\nu} dx \,\frac{x^2 y_\nu(E_x + x)(2k_\nu+n_T+y_{\nu}-E_x)}{E_x (e^x+1)(e^{y_\nu}+1)},
\end{equation}
where $n_T = (\Delta E_{fi}-2m_\nu)/T_{\nu,0}$ is a measure of the kinetic energy of the incident neutrino pair, in units of their temperature, $k_\nu = m_\nu/T_{\nu,0}$ is the ratio of the neutrino mass to the C$\nu$B temperature, and $E_x^2 = k_\nu^2 + x^2$. The remaining variables are
\begin{align}
    x_\nu^+ &= \sqrt{n_T(2k_\nu + n_T)}, \\
    y_\nu &= \sqrt{(2k_\nu + n_T-E_x)^2-k_\nu^2}.
\end{align}
As continuous magnetic fields above $\sim 20\,\mathrm{T}$ are impractical, we limit our search to $m_\nu \lesssim 2\,\mathrm{meV}$. Additionally, as we find that the phase space integral rapidly drops off at above $n_T \simeq 15$, we only compute the phase space integral for $0 \leq n_T \leq 15$. This is somewhat intuitive, as approximately $99.9\%$ of neutrinos have momenta $p_\nu \lesssim 11 \,T_{\nu,0}$, and so the probability of finding a pair of neutrinos with total kinetic energy above $15\,T_{\nu,0}$ is incredibly low.

\begin{figure}[t]
    \includegraphics[width = \linewidth]{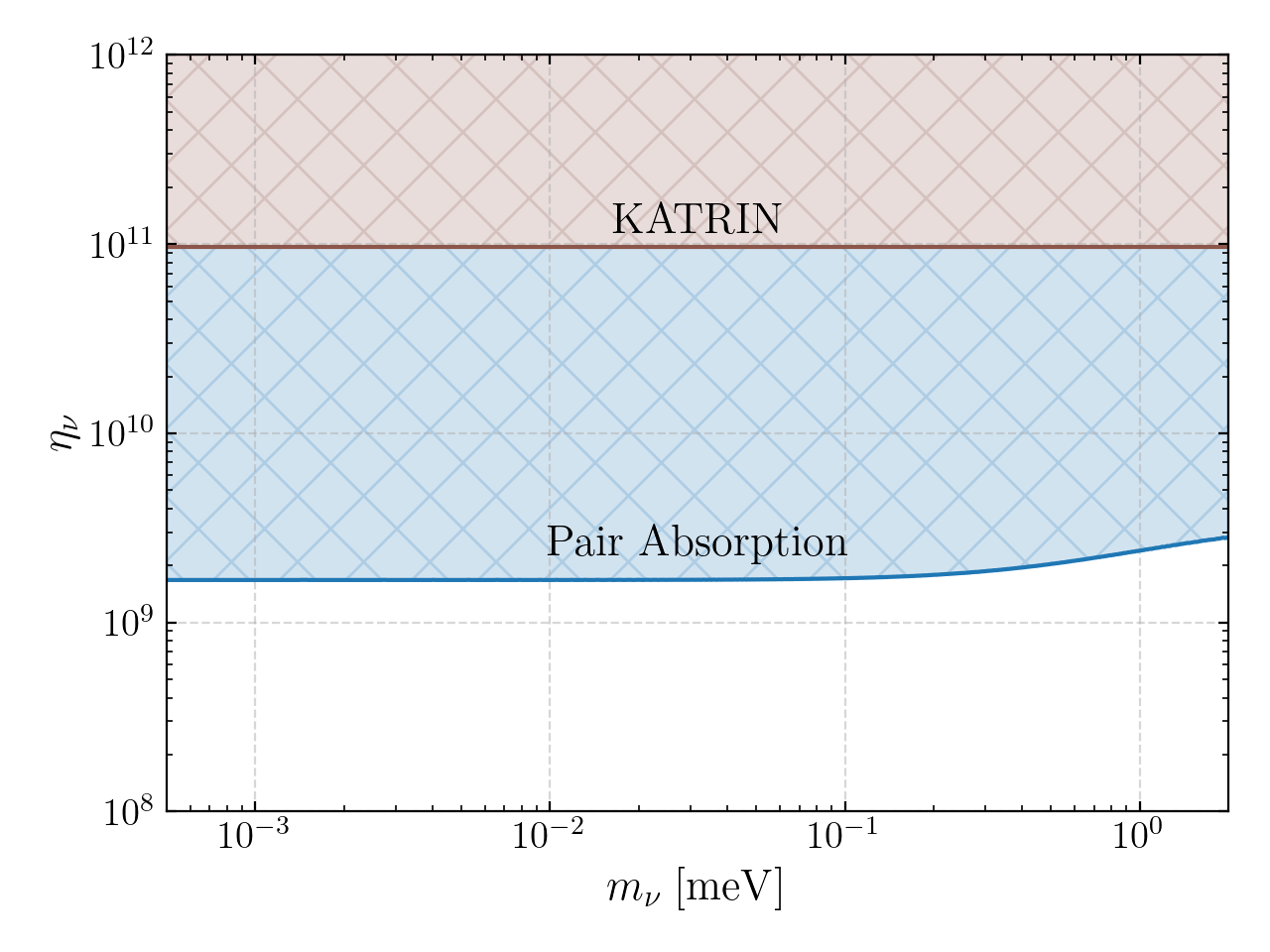}
    \caption{Constraints that could be set on the cosmic neutrino background overdensity parameter, $\eta_\nu$, using spin-flip transitions in hydrogen. We assume an experimental runtime of one year, no Dicke coherence, $10^{3}$ moles of hydrogen as a target, and that a mass range $\Delta m_\nu = 2\,\mathrm{meV}$ is scanned using the applied magnetic field. Also shown is the KATRIN bound from~\cite{KATRIN:2022kkv}.}
    \label{fig:cnub}
\end{figure}

The phase space integral has an exact form in the ultrarelativistic limit as $m_\nu \to 0$, given by
\begin{align}
    \mathcal{P}_{\nu}^\mathrm{UR} &= \frac{96n_T}{e^{n_T}-1}\Bigg(
   \text{Li}_4\left(-e^{n_T}\right)-\frac{\left(n_T^4+7 \pi ^4\right)}{720} \\
   &\hspace{-0.35cm}-\frac{n_T \text{Li}_3\left(-e^{n_T}\right)}{6}- \frac{n_T \zeta (3)}{8}-\frac{2\text{Li}_5\left(-e^{n_T}\right)}{n_T}
   -\frac{15\zeta (5)}{8n_T}\Bigg),\notag
\end{align}
with $\mathrm{Li}_s$ the polylogarithm function of order $s$. This function has a maximum value $\mathcal{P}_{\nu}^\mathrm{UR} \simeq 2.27$ at $n_T \simeq 5.34$, and a width $\Delta n_T \simeq 5.31$. 

For intermediate mass and `heavy' neutrinos, we must instead rely on numerical methods to evaluate integral. Fortunately, as the integral is far simpler to evaluate than those for DM, we can use Gaussian quadrature in place of the methods discussed in Appendix~\ref{app:pysr}.

Now we assume an overdensity in the cosmic neutrino background, \textit{i.e.} we make the replacement $n_{\nu,0} \to n_\nu = \eta_\nu n_{\nu,0}$, with $\eta_\nu$ the overdensity parameter. By inverting~\eqref{eq:cnubRate}, we find that we are able to place  the following constraint on the overdensity
\begin{align}
    \eta_\nu &\lesssim \frac{3\zeta(3)}{\sqrt{\pi \mathcal{P}_\nu}}\frac{\sqrt{T_{\nu,0}}}{G_F\,n_{\nu,0}} \sqrt{\frac{N_\mathrm{events}}{N_T t_\mathrm{eff} }}\\
    &\simeq (1.12 \cdot 10^{9}) \left[\frac{2.27}{\mathcal{P}_\nu}\right]^{\frac{1}{2}}\left[\frac{1\,\mathrm{y}}{t_\mathrm{eff}}\right]^{\frac{1}{2}} \left[\frac{10^{3}}{N_\mathrm{mol}}\right]^{\frac{1}{2}}, \notag
\end{align}
where we assume an observation of $N_\mathrm{events} = 3$, and $N_\mathrm{mol} = 10^{3}$ moles of target hydrogen. The effective exposure time is determined by the mass range scanned, as well as the width of the transition, and is given by
\begin{equation}
    t_\mathrm{eff} = \mathrm{min}\left(1,\frac{\Delta n_T(m_\nu) T_{\nu,0}}{\Delta m_\nu}\right) t_\mathrm{exp},
\end{equation}
with $t_\mathrm{exp}$ the experimental runtime. This width varies from $\Delta n_T \simeq 5.31$ for massless neutrinos, to $\Delta n_T \simeq 1$ at a mass of $2\,\mathrm{meV}$. 

We show the constraints that can be placed on the overdensity parameter, $\eta_\nu$, in Figure~\ref{fig:cnub}. For the mass range accessible using laboratory magnetic fields, pair absorption by hydrogen is able to set constraints approximately $100$ times stronger than KATRIN~\cite{KATRIN:2022kkv}, which currently sets the strongest experimental constraint. For a full discussion of relic neutrino overdensities, and a comparison with other detection proposals, see~\cite{Bauer:2022lri}.

\section{Conclusions}
Constraining light DM via scattering is an incredible challenge due to the tiny momentum transfer in each scattering event. Dark matter absorption overcomes this issue by making use of the full mass-energy of the absorbed particles, leading to much more energetic processes for the same DM mass as scattering. Unlike single particle absorption, however, the operators that give rise to pair absorption do not lead to unstable DM. As an additional benefit, the pair absorption rate scales with the square of the DM number density, $n_\mathrm{DM} \sim 1/m_\mathrm{DM}$, resulting in an enhancement for scalars at very small masses.

We discuss the sensitivity of pair absorption via atomic transitions to light DM, focusing on spin-flip, fine structure, and principal quantum number transitions. All of these processes are forbidden to photons at leading order, and so are largely background free. On the other hand, DM with axial-vector couplings can access all of these transitions at leading order, whilst scalar couplings can only access principal quantum number transitions. 

We demonstrate that sensitivity of atomic transitions to pair absorption of light scalars far exceeds that of scattering. Spin-flip transitions, where the transition energy is set by the strength of an applied magnetic field, are capable of probing electroweak scale axial-vector couplings, $G_{S,A} \simeq G_F$, for a broad range of masses from $\mu\mathrm{eV}$ to meV. Fine structure couplings are capable of setting similarly strong constraints on the axial-vector coupling in the meV to eV range, but only in very narrow mass bands of width neV. Finally, principal quantum number transitions are capable of setting strong constraints on scalar coupling, $G_{S,S} \simeq G_F$, for scalar DM with masses from 1 to 10 eV. Much like fine structure transitions, the mass bands that can be constrained with principal quantum number transitions are very narrow, of order $\mu\mathrm{eV}$. Unfortunately, due to the exclusion principle, light fermionic DM is very hard to constrain with pair absorption, as the density must necessarily be very small at low masses.

We also assess the sensitivity of pair absorption to neutrinos, in particular those from the cosmic neutrino background. Whilst a detection of the cosmic neutrino background via this method is infeasible, we find that relic neutrino pair absorption could set a constraint on the overdensity parameter of $\eta_\nu \lesssim 10^9$ for neutrino masses $m_\nu \lesssim 1\,\mathrm{meV}$, which is roughly two orders of magnitude stronger than the current best constraint, set by KATRIN.

\section*{Acknowledgements}
We would like to thank Martin Hirsch, Alejandro Ibarra, Clare Burrage, Robert McGehee, and Ciaran O'Hare for helpful discussions during the preparation of this manuscript. Martin Bauer is supported by the UKRI Future Leadership Fellowship DARKMAP. Javier Perez-Soler is supported by the grant CIACIF/2022/158, funded by Generalitat Valenciana. This work is further supported by the Spanish grants PID2023-147306NB-I00 and CEX2023-001292-S (MCIU/AEI/10.13039/501100011033), as well as CIPROM/2021/054 (Generalitat Valenciana). 

\appendix
\section{Phase space integrals and transition widths}\label{app:pysr}
In this appendix we give some analytic approximations to the phase space integrals $\mathcal{P}$, $\mathcal{S}_0$, and $\mathcal{S}_Q$, found using the \texttt{PySR} symbolic regression package~\cite{pysr}, and use them to estimate the range of DM masses that can be constrained for a given transition energy. 

To arrive at our analytic approximations, we first evaluate the integrals using the \texttt{VEGAS} algorithm~\cite{PETERLEPAGE1978192} at a set of $10^4$ random $(n,c_\theta)$ grid points, with $n = n_\mathrm{p} \in [0,49/4]$ for pair absorption, $n = n_\mathrm{s} \in [0,49/8]$ for scattering, and $c_\theta \equiv \cos\theta_\mathTerra \in [-1,1]$. For the helicity dependent phase space integrals, we also randomly assign a combination of $h_1$, $h_2 \in \{1,-1\}$ to each grid point. We then fit each of the numerical datasets using the \texttt{PySR} symbolic regression package~\cite{pysr}, limiting our function space to polynomials and the Gaussian defined as
\begin{eqnarray}
    \mathcal{G}_n(\alpha) = \exp\left(-\alpha n^2 \right),
\end{eqnarray}
such that any functions that we obtain are easily integrable and differentiable. The results for each phase space integral are given in Table~\ref{tab:pysr-fits}. For each, we also give the root-mean-square error (RMSE) over the entire parameter space, along with its percentage of the maximum fit value (PMV), defined in turn by
\begin{align}
    \mathrm{RMSE} &= \sqrt{\frac{1}{N}\sum_{i=1}^N\left|f_\mathrm{fit}(\vec{x}_i)-f_\mathrm{data}(\vec{x}_i)\right|^2} \\
    \mathrm{PMV} &= \frac{\mathrm{RMSE}}{\mathrm{max}(f_\mathrm{fit})},
\end{align}
where $\vec{x}_i$ denotes one of the random $(n,c_\theta)$ grid points. We also give the maximum of the fitted function and its coordinates $(n_\mathrm{max},c_{\theta,\mathrm{max}})$. Finally, we give the widths of the phase space integrals in each coordinate, $\Delta n$ and $\Delta c_\theta$, which we estimate as the full width at half maximum of the fitted function. If, at the domain boundary, the function does not reach half of its maximum value, we use the 
boundary point 

The widths, particularly $\Delta n$, are relevant when discussing the constraints on the DM mass that can be placed for a given transition energy. Recall that $n_\mathrm{p}$ and $n_\mathrm{s}$ are defined by
\begin{equation}
    \Delta E_{fi} = \begin{cases}
        2m_\mathrm{DM} + n_\mathrm{p} m_\mathrm{DM} \beta_c^2, &\; \text{pair absorption},\\
        n_\mathrm{s} m_\mathrm{DM} \beta_c^2, &\; \text{scattering},
    \end{cases}
\end{equation}
and effectively serve as measures of the total kinetic energy carried by the absorbed or scattered DM. Now, if the DM were monochromatic, we would only be able to constrain one mass value corresponding a single value of $n$. As this is not the case, we can instead constrain a range of masses corresponding to $n \in [n_\mathrm{max} - \Delta n/2, n_\mathrm{max} + \Delta n/2]$, namely
\begin{equation}\label{eq:mass_width}
    m_\mathrm{DM} \simeq \begin{cases}
        \frac{\Delta E_{fi}}{2 + \beta_c^2\left(n_\mathrm{max} \pm \frac{\Delta n}{2}\right)}, &\quad \text{pair absorption},\\
        \frac{\Delta E_{fi}}{\beta_c^2\left(n_\mathrm{max} \pm \frac{\Delta n}{2}\right)}, &\quad \text{scattering},\\
    \end{cases}
\end{equation}
which has a width comparable with the kinetic energy of the DM in both cases. A consequence of this is that if we want to use a magnetic field to scan over a range of DM masses, $\Delta m$, in the case of spin-flip transitions, then we will need to do so in approximately $\Delta m/\delta m$ steps, with $\delta m$ the range of mass values in~\eqref{eq:mass_width}. It is therefore much easier to constrain couplings via processes with a broader phase space integral, as a sweep over mass space can be performed in fewer steps, allowing us to spend more time on each mass value. 

\section{Transition tables}\label{app:transition-tables}
In this appendix we list the multi-electron atom transitions used in Sections~\ref{sec:fs-transitions} and~\ref{sec:pqn-transitions}. Table~\ref{tab:fs_energies} contains the energy of fine structure transitions from the ground state for all stable, multi-electron atoms. Tables~\ref{tab:principal1} and~\ref{tab:principal2} contain the energy, effective charge, and an estimate of the radial overlap integral for principal quantum number transitions, again considering only ground state transitions for all stable, multi-electron atoms. See the relevant sections for more details. In all cases, we use the levels and energies computed using self consistent field methods, as listed by NIST~\cite{NIST_ASD}.

\begin{table*}[t]
\centering
\renewcommand{\arraystretch}{1.3}
\begin{tabular}{c|c|c|c|c|c}
\hline
Integral & Expression & Max & RMSE (PMV) & $(n_\mathrm{max},c_{\theta,\mathrm{max}})$ & $(\Delta n, \Delta c_\theta)$ \\ \hline\hline
$\mathcal{P}[(\vec{u}-\vec{v})_\perp^2]$ & $(n+3.85)^2 \left(\mathcal{G}_n(0.112)-\mathcal{G}_n(0.557)\right)$ & 19.0 & $0.0629\,(0.332\%)$ & $(2.34,-)$ & $(3.09,-)$ \\ \hline
$\mathcal{S}_0[(\vec{u}+\vec{v})_\perp^2]$ & $0.394 (n-5.45)^3 \left(2.54 c_{\theta }^2+2.54 \mathcal{G}_n(24.1)+n-7.64\right)$ & 381 & $2.95\,(0.777\%)$ & $(0.282,0)$ & $(1.24,-)$ \\ \hline
$\mathcal{S}_Q[(\vec{u}+\vec{v})_\perp^2]$ & $(n-6.23)^2 \left(1.52- c_{\theta }^2\right) \left(\mathcal{G}_n(0.0995)- (n-0.701)^2 \mathcal{G}_n(6.30)\right)$ & 49.1 & $0.322\,(0.657\%)$ & $(0.395,0)$ & $(1.67,1.76)$ \\ \hline
$\mathcal{P}[1]$ & $2.94 \mathcal{G}_n(0.115) \left(-3.90 \mathcal{G}_n(0.633)+n+3.83\right)$ & 10.4 & $0.0355\,(0.341\%)$ & $(1.80,-)$ & $(2.85,-)$ \\ \hline
$\mathcal{S}_0[1]$ & $\mathcal{G}_n(0.277) \left(41.6 n \left(\mathcal{G}_n(1.12)+2.13 \mathcal{G}_n(4.03)\right)+41.5\right)$ & 70.9 & $0.285\,(0.402\%)$ & $(0.380,-)$ & $(1.24,-)$ \\ \hline
$\mathcal{S}_Q[1]$ & $5.89 \left(\mathcal{G}_n(0.243)+n \left(n- (n-1.62)^3\right) \mathcal{G}_n(1.40)\right)$ & 9.85 & $0.0351\,(0.356\%)$ & $(0.353,-)$ & $(1.30,-)$ \\ \hline
\multirow{4}{*}{$\mathcal{P}[f(u,v)]$} & \multirow{4}{*}{$\begin{aligned}&\left(\mathcal{G}_n(0.100)-\mathcal{G}_n(0.445)\right) \left(-h_2 c_{\theta }+\mathcal{G}_n(0.414)+1.41\right)\\ &\times\left(12.8-(n+5.80) \left(h_1 c_{\theta }-\mathcal{G}_n(2.74)\right)\right)\end{aligned}$} & 26.9 & $0.0554\,(0.210\%)$ & $(1.99,1)$ & $(2.97,0.788)$ \\ \cline{3-6} 
 &  & 10.4 & $0.0803\,(0.770\%)$ & $(1.81,0)$ & $(2.85,-)$ \\ \cline{3-6} 
 &  & 10.4 & $0.0756\,(0.725\%)$ & $(1.81,0)$ & $(2.85,-)$ \\ \cline{3-6} 
 &  & 26.9 & $0.0530\,(0.200\%)$ & $(1.99,-1)$ & $(2.97,0.788)$ \\ \hline
\multirow{4}{*}{$\mathcal{S}_0[f(u,v)]$} & \multirow{4}{*}{$\begin{aligned}&\left(\mathcal{G}_n(0.274)+n \left(\mathcal{G}_n(1.37)+2.14 \mathcal{G}_n(4.32)\right)\right)\\ &\times \left(41.4-30.7 c_{\theta } \left(0.413 h_2 \mathcal{G}_n(12.0)+h_1\right)\right)\end{aligned}$} & 128 & $0.972\,(0.783\%)$ & $(0.319,1)$ & $(1.24,1.11)$ \\ \cline{3-6} 
 &  & 120 & $0.865\,(0.741\%)$ & $(0.422,1)$ & $(1.24,1.21)$ \\ \cline{3-6} 
 &  & 120 & $0.946\,(0.809\%)$ & $(0.422,-1)$ & $(1.24,1.21)$ \\ \cline{3-6} 
 &  & 128 & $0.922\,(0.727\%)$ & $(0.319,-1)$ & $(1.24,1.11)$ \\ \hline
\multirow{4}{*}{$\mathcal{S}_Q[f(u,v)]$} & \multirow{4}{*}{$\begin{aligned}&\left(\mathcal{G}_n(0.270)+1.65 n \mathcal{G}_n(2.53)\right) \\ &\times\left(7.00-h_2 c_{\theta } \left(-2.81 h_1 c_{\theta }+h_2 h_1 (n+4.08)+4.03\right)\right)\end{aligned}$} & 25.6 & $0.188\,(0.784\%)$ & $(0.417,1)$ & $(1.36,0.768)$ \\ \cline{3-6} 
 &  & 9.81 & $0.154\,(1.57\%)$ & $(0.402,0.0796)$ & $(1.36,-)$ \\ \cline{3-6} 
 &  & 9.81 & $0.153\,(1.56\%)$ & $(0.402,-0.0796)$ & $(1.36,-)$ \\ \cline{3-6} 
 &  & 25.6 & $0.185\,(0.751\%)$ & $(0.417,-1)$ & $(1.36,0.768)$ \\ \hline
$\mathcal{P}[(\vec{u}-\vec{v})^2]$ & $2.15\, \mathcal{G}_n(0.0912) \left(32.8-\left((n-0.957)^3+33.5\right) \mathcal{G}_n(0.217)\right)$ & 28.4 & $0.0641\,(0.226\%)$ & $(2.35,-)$ & $(3.09,-)$ \\ \hline
$\mathcal{S}_0[(\vec{u}+\vec{v})^2]$ & $11.7 \mathcal{G}_n(0.234) \left(\mathcal{G}_n(0.880) \left(19.8-10.7 \mathcal{G}_n(45.1)\right)+24.6\right)$ & 494 & $2.65\,(0.536\%)$ & $(0.252,-)$ & $(1.24,-)$ \\ \hline
$\mathcal{S}_Q[(\vec{u}+\vec{v})^2]$ & $\mathcal{G}_n(0.302) \left((n-0.741)^3 \left(59.4 \mathcal{G}_n(10.3)+4.85\right)+60.3\right)$ & 56.9 & $0.239\,(0.420\%)$ & $(0.373,-)$ & $(1.67,-)$ \\ \hline
\multirow{4}{*}{$\mathcal{P}[g(u,v)]$} & \multirow{4}{*}{$\begin{aligned}&\left(\mathcal{G}_n(0.0990)-\mathcal{G}_n(0.582)\right)\\
&\times\left(30.0 \mathcal{G}_n(1.06)+1.38 h_1 h_2 (n+2.86)+53.1\right)\end{aligned}$} & 34.8 & $0.0627\,(0.180\%)$ & $(1.89,-)$ & $(2.85,-)$ \\ \cline{3-6} 
 &  & 27.3 & $0.0816\,(0.298\%)$ & $(1.79,-)$ & $(2.85,-)$ \\ \cline{3-6} 
 &  & 27.3 & $0.0806\,(0.295\%)$ & $(1.79,-)$ & $(2.85,-)$ \\ \cline{3-6} 
 &  & 34.8 & $0.0625\,(0.179\%)$ & $(1.89,-)$ & $(2.85,-)$ \\ \hline
\multirow{4}{*}{$\mathcal{S}_0[g(u,v)]$} & \multirow{4}{*}{$\begin{aligned}&\mathcal{G}_n(5.34) \left(8.4 h_1 h_2+303 n-89\right)\\&+109 \mathcal{G}_n(0.259)+109 \mathcal{G}_n(0.918)\end{aligned}$} & 217 & $0.839\,(0.386\%)$ & $(0.368,-)$ & $(1.24,-)$ \\ \cline{3-6} 
 &  & 210 & $0.815\,(0.389\%)$ & $(0.397,-)$ & $(1.24,-)$ \\ \cline{3-6} 
 &  & 210 & $0.919\,(0.439\%)$ & $(0.397,-)$ & $(1.24,-)$ \\ \cline{3-6} 
 &  & 217 & $0.890\,(0.410\%)$ & $(0.368,-)$ & $(1.24,-)$ \\ \hline
\multirow{4}{*}{$\mathcal{S}_Q[g(u,v)]$} & \multirow{4}{*}{$\begin{aligned}&\mathcal{G}_n(0.343) \Big[-\left((2.55-1.90 n)^3-19.1\right) \\
&\times\left(\mathcal{G}_n(2.73)+0.597\right)+3.74 h_1 h_2+14.5\Big]\end{aligned}$} & 33.2 & $0.209\,(0.630\%)$ & $(0.358,-)$ & $(1.36,-)$ \\ \cline{3-6} 
 &  & 26.0 & $0.174\,(0.667\%)$ & $(0.373,-)$ & $(1.36,-)$ \\ \cline{3-6} 
 &  & 26.0 & $0.175\,(0.673\%)$ & $(0.373,-)$ & $(1.36,-)$ \\ \cline{3-6} 
 &  & 33.2 & $0.191\,(0.576\%)$ & $(0.358,-)$ & $(1.36,-)$ \\ \hline
\end{tabular}
\caption{Analytic approximations to the phase space integrals found using the \texttt{PySR} symbolic regression package~\cite{pysr}. For each fit, we list the maximum value of the function, the root-mean-square error (RMSE) and its value as a percentage of the fit maximum. We also give the coordinates of the maximum value in the $(n,c_\theta)$ plane, along with the width of the transition in both variable $(\Delta n, \Delta c_\theta)$. The functions $f$ and $g$ refer to the bracketed terms in~\eqref{eq:fermion-spin-flip-amp} and~\eqref{eq:amp-fs-f}, and from top to bottom for each, the four sets of values correspond, in turn, to $(h_1,h_2) = (-1,-1)$, $(-1,1)$, $(1,-1)$, and $(1,1)$. See the text for more details.}
\label{tab:pysr-fits}
\end{table*}

\begin{table*}[b]
\centering
\renewcommand{\arraystretch}{1.15}
\begin{tabular}{ccc}
\hline
Element & Transition & $\Delta E_{fi}\;[\mathrm{meV}]$ \\
\hline\hline
B  & $^{2}P^{\circ}_{1/2}\to{}^{2}P^{\circ}_{3/2}$ & $1.895$   \\
C  & $^{3}P_{0}\to{}^{3}P_{1}$             & $2.035$   \\
O  & $^{3}P_{2}\to{}^{3}P_{1}$             & $19.62$   \\
F  & $^{2}P^{\circ}_{3/2}\to{}^{2}P^{\circ}_{1/2}$ & $50.11$  \\
\hline\hline
Al & $^{2}P^{\circ}_{1/2}\to{}^{2}P^{\circ}_{3/2}$ & $13.89$  \\
Si & $^{3}P_{0}\to{}^{3}P_{1}$             & $9.561$  \\
S  & $^{3}P_{2}\to{}^{3}P_{1}$             & $49.10$  \\
Cl & $^{2}P^{\circ}_{3/2}\to{}^{2}P^{\circ}_{1/2}$ & $109.4$  \\
\hline\hline
Sc & $^{2}D_{3/2}\to{}^{2}D_{5/2}$         & $20.87$  \\
Ti & $^{3}F_{2}\to{}^{3}F_{3}$             & $21.09$  \\
V  & $^{4}F_{3/2}\to{}^{4}F_{5/2}$         & $17.03$  \\
Fe & $^{5}D_{4}\to{}^{5}D_{3}$             & $51.57$  \\
Co & $^{4}F_{9/2}\to{}^{4}F_{7/2}$         & $101.2$  \\
Ni & $^{3}F_{4}\to{}^{3}F_{3}$             & $165.2$  \\
\hline
Ga & $^{2}P^{\circ}_{1/2}\to{}^{2}P^{\circ}_{3/2}$ & $102.4$ \\
Ge & $^{3}P_{0}\to{}^{3}P_{1}$             & $69.08$  \\
Se & $^{3}P_{2}\to{}^{3}P_{1}$             & $246.7$  \\
Br & $^{2}P^{\circ}_{3/2}\to{}^{2}P^{\circ}_{1/2}$ & $456.9$ \\
\hline\hline
Y  & $^{2}D_{3/2}\to{}^{2}D_{5/2}$         & $65.76$  \\
Zr & $^{3}F_{2}\to{}^{3}F_{3}$             & $70.72$  \\
Nb & $^{6}D_{1/2}\to{}^{6}D_{3/2}$         & $19.12$  \\
Ru & $^{5}F_{5}\to{}^{5}F_{4}$             & $147.6$  \\
Rh & $^{4}F_{9/2}\to{}^{4}F_{7/2}$         & $189.7$  \\
\hline
\end{tabular}
\qquad
\begin{tabular}{ccc}
\hline
Element & Transition & $\Delta E_{fi}\;[\mathrm{meV}]$ \\
\hline\hline
In  & $^{2}P^{\circ}_{1/2}\to{}^{2}P^{\circ}_{3/2}$ & $274.3$  \\
Sn  & $^{3}P_{0}\to{}^{3}P_{1}$             & $209.8$  \\
Te  & $^{3}P_{2}\to{}^{3}P_{0}$             & $583.5$  \\
I   & $^{2}P^{\circ}_{3/2}\to{}^{2}P^{\circ}_{1/2}$ & $942.6$  \\
\hline\hline
La  & $^{2}D_{3/2}\to{}^{2}D_{5/2}$         & $130.5$  \\
Pr  & $^{4}I^{\circ}_{9/2}\to{}^{4}I^{\circ}_{11/2}$ & $170.7$ \\
Nd  & $^{5}I_{4}\to{}^{5}I_{5}$             & $139.9$  \\
Sm  & $^{7}F_{0}\to{}^{7}F_{1}$             & $36.27$  \\
Gd  & $^{9}D^{\circ}_{2}\to{}^{9}D^{\circ}_{3}$ & $26.67$ \\
Tb  & $^{6}H^{\circ}_{15/2}\to{}^{6}H^{\circ}_{13/2}$ & $343.6$ \\
Dy  & $^{5}I_{8}\to{}^{5}I_{7}$             & $512.6$  \\
Ho  & $^{4}I^{\circ}_{15/2}\to{}^{4}I^{\circ}_{13/2}$ & $672.0$ \\
Er  & $^{3}H_{6}\to{}^{3}H_{5}$             & $862.7$  \\
Tm  & $^{2}F^{\circ}_{7/2}\to{}^{2}F^{\circ}_{5/2}$ & $1087$  \\
Lu  & $^{2}D_{3/2}\to{}^{2}D_{5/2}$         & $247.2$  \\
\hline
Hf  & $^{3}F_{2}\to{}^{3}F_{3}$             & $292.2$  \\
Ta  & $^{4}F_{3/2}\to{}^{4}F_{5/2}$         & $249.2$  \\
W   & $^{5}D_{0}\to{}^{5}D_{1}$             & $207.1$  \\
Os  & $^{5}D_{4}\to{}^{5}D_{3}$             & $515.7$  \\
Ir  & $^{4}F_{9/2}\to{}^{4}F_{7/2}$         & $784.1$  \\
Pt  & $^{3}D_{3}\to{}^{3}D_{2}$             & $814.3$  \\
\hline
Tl  & $^{2}P^{\circ}_{1/2}\to{}^{2}P^{\circ}_{3/2}$ & $966.2$ \\
Pb  & $^{3}P_{0}\to{}^{3}P_{1}$             & $969.5$  \\
\hline\hline
\end{tabular}
\caption{Fine structure transition energies for all stable, multi-electron atoms whose ground state has a fine structure splitting, separated by block and period. Energies and levels taken from NIST~\cite{NIST_ASD}.}
\label{tab:fs_energies}
\end{table*}

\begin{table*}[t]
\centering
\renewcommand{\arraystretch}{1.15}
\begin{tabular}{ccccc}
\hline
Element & Term (Electron) & $\Delta E_{fi}\;[\mathrm{eV}]$ & $Z_\mathrm{eff}$ & $\mathcal{I}_{gg} \simeq -\mathcal{I}_{ff}\;[10^{-5}]$ \\
\hline\hline
H  & $^{2}S_{1/2}\,(1s_{1/2} \to 2s_{1/2})$ & $10.20$ & $1$        & $0.55785$     \\
He & $^{1}S_{0}\,(1s_{1/2} \to 2s_{1/2})$ & $20.62$ & $1.47(22)$ & $1.29(40)$    \\
\hline\hline
Li & $^{2}S_{1/2}\,(2s_{1/2} \to 3s_{1/2})$ & $3.373$ & $1.39(10)$ & $0.380(58)$   \\
Be & $^{1}S_{0}\,(2s_{1/2} \to 3s_{1/2})$ & $6.779$ & $1.90(29)$ & $0.70(22)$  \\
\hline
B  & $^{2}P^{\circ}_{1/2}\,(2p_{1/2} \to 3p_{1/2})$ & $6.027$  & $2.11(58)$ & $0.76(38)$    \\
C  & $^{3}P_{0}\,(2p_{1/2} \to 3p_{1/2})$ & $8.847$  & $2.54(79)$ & $1.16(64)$    \\
N  & $^{4}S^{\circ}_{3/2}\,(2p_{3/2} \to 3p_{3/2})$ & $12.00$  & $2.9(11)$  & $1.54(99)$    \\
O  & $^{3}P_{2}\,(2p_{1/2} \to 3p_{1/2})$ & $10.99$  & $3.4(13)$  & $2.1(14)$     \\
F  & $^{2}P^{\circ}_{3/2}\,(2p_{3/2} \to 3p_{3/2})$ & $14.75$  & $3.8(15)$  & $2.8(19)$     \\
Ne & $^{1}S_{0}\,(2p_{3/2} \to 3p_{3/2})$ & $18.71$  & $4.2(17)$  & $3.4(24)$     \\
\hline\hline
Na & $^{2}S_{1/2}\,(3s_{1/2} \to 4s_{1/2})$ & $3.191$ & $3.07(52)$ & $0.93(30)$    \\
Mg & $^{1}S_{0}\,(3s_{1/2} \to 4s_{1/2})$ & $5.393$ & $3.59(69)$ & $1.32(52)$    \\
\hline
Al & $^{2}P^{\circ}_{1/2}\,(3p_{1/2} \to 4p_{1/2})$ & $4.085$ & $3.10(90)$ & $0.98(50)$    \\
Si & $^{3}P_{0}\,(3p_{1/2} \to 4p_{1/2})$ & $6.079$ & $3.6(11)$& $1.34(77)$    \\
P  & $^{4}S^{\circ}_{3/2}\,(3p_{3/2} \to 4p_{3/2})$ & $8.286$ & $4.0(13)$  & $1.5(10)$     \\
S  & $^{3}P_{2}\,(3p_{1/2} \to 4p_{1/2})$ & $8.046$ & $4.5(15)$  & $2.1(14)$     \\
Cl & $^{2}P^{\circ}_{3/2}\,(3p_{3/2} \to 4p_{3/2})$ & $10.59$ & $5.0(17)$  & $2.5(16)$     \\
Ar & $^{1}S_{0}\,(3p_{3/2} \to 4p_{3/2})$ & $13.27$ & $5.5(19)$  & $3.1(19)$     \\
\hline\hline
K  & $^{2}S_{1/2}\,(4s_{1/2} \to 5s_{1/2})$ & $2.607$ & $4.58(65)$ & $1.24(34)$    \\
Ca & $^{1}S_{0}\,(4s_{1/2} \to 5s_{1/2})$ & $4.131$ & $5.12(81)$ & $1.55(47)$    \\
\hline
Sc & $^{2}D_{3/2}\,(4s_{1/2} \to 5s_{1/2})$ & $4.423$ & $5.37(87)$ & $1.72(55)$    \\
Ti & $^{3}F_{2}\,(4s_{1/2} \to 5s_{1/2})$ & $4.654$ & $5.63(92)$ & $1.87(57)$    \\
V  & $^{4}F_{3/2}\,(4s_{1/2} \to 5s_{1/2})$ & $4.851$ & $5.88(97)$ & $2.12(63)$    \\
Cr & $^{7}S_{3}\,(4s_{1/2} \to 5s_{1/2})$ & $4.574$ & $5.85(92)$ & $2.05(63)$    \\
Mn & $^{6}S_{5/2}\,(4s_{1/2} \to 5s_{1/2})$ & $5.133$ & $6.4(11)$  & $2.45(80)$    \\
Fe & $^{5}D_{4}\,(4s_{1/2} \to 5s_{1/2})$ & $5.539$ & $6.7(11)$  & $2.66(90)$    \\
Co & $^{4}F_{9/2}\,(4s_{1/2} \to 5s_{1/2})$ & $5.892$ & $6.9(12)$  & $2.85(99)$    \\
Ni & $^{3}F_{4}\,(4s_{1/2} \to 5s_{1/2})$ & $6.257$ & $7.2(12)$  & $3.1(10)$   \\
Cu & $^{2}S_{1/2}\,(4s_{1/2} \to 5s_{1/2})$ & $5.348$ & $7.1(12)$  & $2.9(10)$   \\
Zn & $^{1}S_{0}\,(4s_{1/2} \to 5s_{1/2})$ & $6.917$ & $7.7(13)$  & $3.4(12)$   \\
\hline
Ga & $^{2}P^{\circ}_{1/2}\,(4p_{1/2} \to 5p_{1/2})$ & $4.096$ & $6.3(12)$  & $2.32(77)$    \\
Ge & $^{3}P_{0}\,(4p_{1/2} \to 5p_{1/2})$ & $5.890$ & $6.8(14)$  & $2.8(11)$   \\
As & $^{4}S^{\circ}_{3/2}\,(4p_{3/2} \to 5p_{3/2})$ & $7.891$ & $7.3(16)$  & $3.2(13)$   \\
Se & $^{3}P_{2}\,(4p_{1/2} \to 5p_{1/2})$ & $7.523$ & $7.8(17)$  & $3.7(15)$   \\
Br & $^{2}P^{\circ}_{3/2}\,(4p_{3/2} \to 5p_{3/2})$ & $9.881$ & $8.3(19)$  & $4.1(18)$   \\
Kr & $^{1}S_{0}\,(4p_{3/2} \to 5p_{3/2})$ & $11.66$ & $8.8(21)$  & $4.5(22)$   \\
\hline\hline
\end{tabular}
\caption{Principal quantum number transition parameters for all stable elements up to $Z = 36$, separated by block and period. Energies taken from~\cite{NIST_ASD}, and $Z_\mathrm{eff}$ estimated using data from~\cite{LANZINI2015240}. See the text in Section~\ref{sec:pqn-transitions} for more details.}
\label{tab:principal1}
\end{table*}

\begin{table*}[t]
\centering
\renewcommand{\arraystretch}{1.15}
\begin{tabular}{ccccc}
\hline
Element & Term (Electron) & $\Delta E_{fi}\;[\mathrm{eV}]$ & $Z_\mathrm{eff}$ & $\mathcal{I}_{gg}\simeq -\mathcal{I}_{ff}\;[10^{-5}]$ \\
\hline\hline
Rb & $^{2}S_{1/2}\,(5s_{1/2} \to 6s_{1/2})$ & $2.496$ & $7.8(12)$ & $2.42(70)$ \\
Sr & $^{1}S_{0}\,(5s_{1/2} \to 6s_{1/2})$ & $3.793$ & $8.3(13)$ & $2.72(81)$ \\
\hline
Y  & $^{2}D_{3/2}\,(5s_{1/2} \to 6s_{1/2})$ & $4.515$ & $8.6(14)$ & $3.03(84)$ \\
Zr & $^{3}F_{2}\,(5s_{1/2} \to 6s_{1/2})$ & $4.644$ & $8.9(14)$ & $3.15(97)$ \\
Nb & $^{6}D_{1/2}\,(5s_{1/2} \to 6s_{1/2})$ & $4.638$ & $9.0(14)$ & $3.3(10)$ \\
Mo & $^{7}S_{3}\,(5s_{1/2} \to 6s_{1/2})$ & $4.919$ & $9.3(15)$ & $3.3(11)$ \\
Ru & $^{5}F_{5}\,(5s_{1/2} \to 6s_{1/2})$ & $5.115$ & $9.9(16)$ & $3.9(12)$ \\
Rh & $^{4}F_{9/2}\,(5s_{1/2} \to 6s_{1/2})$ & $5.192$ & $10.2(17)$& $4.1(13)$ \\
Pd & $^{1}S_{0}\,(4d_{5/2} \to 5d_{5/2})$ & $6.865$ & $12.2(45)$& $8.1(54)$ \\
Ag & $^{2}S_{1/2}\,(5s_{1/2} \to 6s_{1/2})$ & $5.276$ & $10.8(18)$& $4.5(15)$ \\
Cd & $^{1}S_{0}\,(5s_{1/2} \to 6s_{1/2})$ & $6.610$ & $11.4(19)$& $5.2(16)$ \\ \hline
In  & $^{2}P^{\circ}_{1/2}\,(5p_{1/2} \to 6p_{1/2})$ & $3.945$ & $9.8(18)$  & $3.8(14)$   \\
Sn  & $^{3}P_{0}\,(5p_{1/2} \to 6p_{1/2})$ & $5.430$ & $10.3(20)$ & $4.0(15)$   \\
Sb  & $^{4}S^{\circ}_{3/2}\,(5p_{3/2} \to 6p_{3/2})$ & $6.523$ & $10.8(22)$ & $4.6(18)$   \\
Te  & $^{3}P_{2}\,(5p_{1/2} \to 6p_{1/2})$ & $6.902$ & $11.4(23)$ & $5.0(20)$   \\
I   & $^{2}P^{\circ}_{3/2}\,(5p_{3/2} \to 6p_{3/2})$ & $8.924$ & $11.8(25)$ & $5.5(22)$   \\
Xe  & $^{1}S_{0}\,(5p_{3/2} \to 6p_{3/2})$ & $9.933$ & $12.4(27)$ & $6.2(25)$   \\
\hline\hline
Cs  & $^{2}S_{1/2}\,(6s_{1/2} \to 7s_{1/2})$ & $2.298$ & $10.0(22)$ & $3.1(12)$   \\
Ba  & $^{1}S_{0}\,(6s_{1/2} \to 7s_{1/2})$ & $3.500$ & $10.6(23)$ & $3.3(15)$   \\
\hline
La  & $^{2}D_{3/2}\,(6s_{1/2} \to 7s_{1/2})$ & $3.929$ & $10.9(24)$ & $3.6(14)$   \\
Sm  & $^{7}F_{0}\,(6s_{1/2} \to 7s_{1/2})$ & $3.653$ & $11.7(26)$ & $4.1(16)$   \\
Eu  & $^{8}S^{\circ}_{7/2}\,(6s_{1/2} \to 7s_{1/2})$ & $3.659$ & $11.9(27)$ & $4.1(18)$   \\
Gd  & $^{9}D^{\circ}_{2}\,(6s_{1/2} \to 7s_{1/2})$ & $4.086$ & $12.2(28)$ & $4.3(19)$   \\
Tb  & $^{6}H^{\circ}_{15/2}\,(6s_{1/2} \to 7s_{1/2})$ & $3.794$ & $12.2(27)$ & $4.5(17)$   \\
Dy  & $^{5}I_{8}\,(6s_{1/2} \to 7s_{1/2})$ & $4.330$ & $12.4(28)$ & $4.6(20)$   \\
Ho  & $^{4}I^{\circ}_{15/2}\,(6s_{1/2} \to 7s_{1/2})$ & $4.092$ & $12.6(28)$ & $4.6(19)$   \\
Er  & $^{3}H_{6}\,(6s_{1/2} \to 7s_{1/2})$ & $4.156$ & $12.8(29)$ & $4.7(20)$   \\
Tm  & $^{2}F^{\circ}_{7/2}\,(6s_{1/2} \to 7s_{1/2})$ & $4.211$ & $13.0(29)$ & $4.9(20)$   \\
Yb  & $^{1}S_{0}\,(6s_{1/2} \to 7s_{1/2})$ & $4.259$ & $13.2(30)$ & $5.1(20)$   \\
Lu  & $^{2}D_{3/2}\,(5d_{3/2} \to 6d_{3/2})$ & $3.911$ & $16.2(20)$ & $9.7(24)$   \\
\hline
Hf  & $^{3}F_{2}\,(6s_{1/2} \to 7s_{1/2})$ & $5.120$ & $13.8(32)$ & $5.9(26)$   \\
Ta  & $^{4}F_{3/2}\,(6s_{1/2} \to 7s_{1/2})$ & $5.450$ & $14.2(33)$ & $5.8(27)$   \\
W   & $^{5}D_{0}\,(6s_{1/2} \to 7s_{1/2})$ & $5.607$ & $14.5(34)$ & $6.1(28)$   \\
Re  & $^{6}S_{5/2}\,(6s_{1/2} \to 7s_{1/2})$ & $5.542$ & $14.9(34)$ & $6.4(28)$   \\
Os  & $^{5}D_{4}\,(6s_{1/2} \to 7s_{1/2})$ & $6.328$ & $15.2(35)$ & $6.8(33)$   \\
Pt  & $^{3}D_{3}\,(6s_{1/2} \to 7s_{1/2})$ & $6.494$ & $15.7(37)$ & $7.4(38)$   \\
Au  & $^{2}S_{1/2}\,(6s_{1/2} \to 7s_{1/2})$ & $6.755$ & $16.0(38)$ & $7.4(35)$   \\
Hg  & $^{1}S_{0}\,(6s_{1/2} \to 7s_{1/2})$ & $7.926$ & $16.5(39)$ & $8.1(33)$   \\
\hline
Tl  & $^{2}P^{\circ}_{1/2}\,(6p_{1/2} \to 7p_{1/2})$ & $4.235$ & $13.9(41)$ & $5.9(31)$   \\
Pb  & $^{3}P_{0}\,(6p_{1/2} \to 7p_{1/2})$ & $5.505$ & $14.4(43)$ & $6.0(32)$   \\
\hline\hline
\end{tabular}
\caption{Principal quantum number transition parameters for all stable elements from $Z = 37$ to $Z = 82$, separated by block
and period. Missing are Ce, Pr, Nd, and Ir, for which there are no viable excited states listed by NIST. Energies taken from~\cite{NIST_ASD}, and $Z_\mathrm{eff}$ estimated using data from~\cite{LANZINI2015240}. See the text in Section~\ref{sec:pqn-transitions} for more details.}
\label{tab:principal2}
\end{table*}

\bibliography{bibliography}

\end{document}